\documentclass[trackchanges]{aastex701}

\begin{document}

\title{Antarctic Infrared Binocular Telescope. I.\\ System Overview, Laboratory Testing, and On-Sky Performance Evaluation}

\author[orcid=0009-0003-5592-3734]{Zhongnan Dong}
\affiliation{School of Physics and Astronomy, Sun Yat-sen University, Zhuhai 519082, China}
\email{dongzhongnan@mail.sysu.edu.cn}

\author[orcid=0000-0002-6077-6287]{Bin Ma}
\affiliation{School of Physics and Astronomy, Sun Yat-sen University, Zhuhai 519082, China}
\affiliation{CSST Science Center for the Guangdong-Hong Kong-Macau Greater Bay Area, Zhuhai 519082, China}
\email{mabin3@mail.sysu.edu.cn}
\correspondingauthor{Bin Ma}

\author{Haoran Zhang}
\affiliation{School of Physics and Astronomy, Sun Yat-sen University, Zhuhai 519082, China}
\email{zhanghr33@mail2.sysu.edu.cn}

\author{Jinji Li}
\affiliation{School of Physics and Astronomy, Sun Yat-sen University, Zhuhai 519082, China}
\email{liij328@mail2.sysu.edu.cn}

\author{Xu Yang}
\affiliation{National Astronomical Observatories, Chinese Academy of Sciences, Beijing 100101, China}
\email{xuyang@nao.cas.cn}

\author{Yi Hu}
\affiliation{National Astronomical Observatories, Chinese Academy of Sciences, Beijing 100101, China}
\email{huyi.naoc@gmail.com}

\author{Zhaohui Shang}
\affiliation{National Astronomical Observatories, Chinese Academy of Sciences, Beijing 100101, China}
\email{zshang@gmail.com}

\author{Michael C.~B. Ashley}
\affiliation{School of Physics, University of New South Wales, Sydney, NSW 2052, Australia}
\email{m.ashley@unsw.edu.au}

\collaboration{all}{The Terra Mater collaboration}


\begin{abstract}

Infrared time-domain surveys remain significantly underdeveloped compared with their optical counterparts. We have developed the Antarctic Infrared Binocular Telescope (AIRBT) to study the dynamic infrared sky at Dome A, Antarctica, taking advantage of the superb infrared observational conditions at this site. AIRBT consists of two identical 15\,cm $f/3$ optical tube assemblies and two cost-effective indium gallium arsenide (InGaAs) cameras equipped with $J$ and $H$ filters, respectively. The cameras have 640 $\times$ 512 15~\textmu m pixels, giving a scale of 6.9 arcsec\,pixel$^{-1}$ and a field of view (FoV) of 1.22 $\times$ 0.97 deg$^2$. We characterize the performance of the InGaAs cameras, including bias, readout noise, dark current, nonlinearity, and photon transfer curve (PTC). Our analysis highlights the distinct behaviors of InGaAs cameras compared with charge-coupled devices (CCDs). The bias and readout noise show temperature dependence. In addition, the noise measured from the PTCs has additional components increasing with exposure time. On-sky tests were conducted in October~2022 including system calibration, limiting depth, and photometric precision. For a single 3\,s exposure, we achieved 5$\sigma$ limiting magnitudes of 11.2 mag (Vega system) in $J$ band and 9.7 mag in $H$ band. The best photometric precision reached 20\,mmag at the bright end, which could be further improved to sub-percent levels through image stacking. AIRBT was installed at Dome~A in January~2023, and scientific observations began as soon as darkness set in.

\end{abstract}


\section{Introduction}
\label{sec:introduction}

Infrared surveys are essential for revealing cold, obscured, and/or high-redshift objects such as exoplanets, cold stars, star-forming regions, and distant galaxies. Static infrared surveys have mapped the sky with remarkable area and depth, including the Two Micron All-Sky Survey \citep[2MASS,][]{2006AJ....131.1163S}, the Deep Near-Infrared Southern Sky Survey \citep[DENIS,][]{1997Msngr..87...27E}, the UKIRT Infrared Deep Sky Survey \citep[UKIDSS,][]{2007MNRAS.379.1599L}, and large public surveys conducted by VISTA \citep[]{2015A&A...575A..25S}. 
However, the dynamic infrared sky remains largely unexplored compared to the densely instrumented optical time-domain surveys. 
VISTA allocated fractional observation time for a variability survey of the Milky Way over an area of 520\,deg\textsuperscript{2}, called VISTA Variables in the Via Lactea \citep[VVV,][]{2010NewA...15..433M}. 
The Palomar Gattini-IR \citep[]{2016SPIE.9906E..2CM, 2020PASP..132b5001D, 2024PASP..136j4501M} is a dedicated time-domain infrared survey with a 30\,cm aperture telescope, covering about 15,000 square degrees of accessible sky with a median cadence of 2 days to a depth of 14.9 mag in $J$ band. All magnitudes quoted in this paper are on the Vega system.

Infrared time-domain surveys face several challenges. One of the main issues is the high cost and complex cooling requirements (down to 80\,K) of traditional mercury cadmium telluride (HgCdTe) detectors. Recently, cost-effective indium gallium arsenide (InGaAs) detectors have improved and offer an alternative solution for wavelengths of 0.9\,--\,1.7\,\textmu m. These detectors typically operate at temperatures higher than \textminus80\textcelsius, achievable with simple thermoelectric cooling (TEC) without the need for complex cryogenic systems. Although their dark current is orders of magnitude higher than that of HgCdTe detectors or charge-coupled devices (CCDs), it is comparable to or even smaller than the infrared sky background. Astronomers have begun testing InGaAs cameras in astronomical observations.

\cite{2013PASP..125.1021S} used a 0.6\,m telescope equipped with an InGaAs camera, and achieved sub-mmag photometric precision. Similarly, tests with 12-inch and 18-inch telescopes showed that sub-percent precision could be obtained during exoplanet transits \citep{2018PASP..130i5001S}. An InGaAs camera on a 2.5\,m telescope achieved background-limited imaging, enabling the detection of two supernovae and a 1.2\% transit event \citep{2019AJ....157...46S}. On the 2\,m Liverpool Telescope with an $H$-band filter, an InGaAs camera reached mmag precision for sources $<$ 10.7 mag and a 10$\sigma$ depth of 16 mag with a total exposure time of 3 minutes \citep{2022PASP..134f5001B}.

After extensive on-sky testing, InGaAs cameras are now being deployed in time-domain surveys. 
The Wide-Field Infrared Transient Explorer \citep[WINTER,][]{2020SPIE11447E..9KL} began operations at Palomar Observatory in June 2023, with six 1920 $\times$ 1080 pixel InGaAs cameras on a 1\,m telescope with a field of view (FoV) of 1\,deg\textsuperscript{2}. It is designed for dedicated near-infrared follow-up of kilonovae detected by LIGO \citep{2022ApJ...926..152F}. 
The Dynamic REd All-sky Monitoring Survey \citep[DREAMS,][]{2020SPIE11203E..07S,2022JATIS...8a6001B} is a time-domain sky survey using six 1280 $\times$ 1024 pixel InGaAs cameras on a 0.5\,m telescope at Siding Spring Observatory; commissioning is planned for 2026.
A 1280 $\times$ 1024 pixel InGaAs-based instrument (0.81\,--\,1.33\,\textmu m) was designed for the SPECULOOS telescope \citep{2024SPIE13096E..3XP}. It achieved better photometric precision than CCDs (0.7\,--\,1.1\,\textmu m) for L-type stars and cooler, as the infrared band significantly reduced noise from atmospheric precipitable water vapor (PWV) variability.

Another challenge is that ground-based observations are limited by site conditions. The infrared sky background from 1\,--\,4\,\textmu m at temperate-latitude observatories is typically between 10 and $10^4$ times brighter than in optical bands, due to strong hydroxyl airglow and thermal emission from the atmosphere and telescope \citep[e.g.,][]{2012A&A...543A..92N}. Moreover, atmospheric transmittance is significantly reduced at certain wavelengths due to water vapor absorption, creating gaps between the $J$, $H$, and $K$ bands.
All of these issues are ameliorated when observing from the Antarctic plateau, which is an exceptional site for infrared observations \citep{2000PASA...17..260H}, primarily due to its unique environmental conditions. The extremely low temperatures in Antarctica greatly reduce thermal radiation from the lower atmosphere and the telescope, as well as reducing water vapor absorption since much of the water has precipitated out.
Observations at the South Pole demonstrate a 20\,--\,100 times reduction in background emission in the $K_{\mathrm{dark}}$ band (2.43 \textmu m), along with 2\,--\,3 times reductions in the $J$ and $H$ bands \citep{1996PASP..108..721A, 1999ApJ...527.1009P}.

The excellent infrared sky conditions in Antarctica have led to a number of projects. For example, the 0.6\,m SPIREX telescope at the South Pole Station \citep{2000ApJ...542..359B};
the International Robotic Antarctic Infrared Telescope \citep[IRAIT,][]{2006SPIE.6267E..1HT, 2014SPIE.9145E..0DD} with the near/mid-infrared camera AMICA (1.25\,--\,25\,\textmu m) at Dome C, Antarctica; the proposed Cryoscope telescope
\citet{2025PASP..137f5001K}, a 50\,deg\textsuperscript{2} FoV and 1.2\,m aperture survey telescope in the $K_{\mathrm{dark}}$ band planned for deployment to Dome C, and preceded by a pathfinder prototype with 16\,deg\textsuperscript{2} FoV and 26\,cm aperture scheduled for installation in December 2026 \citep{2024SPIE13096E..3ME}.

Dome A, as the highest location on the Antarctic plateau, has the best published THz and optical observation conditions on the ground \citep{2020RAA....20..168S}. It has the lowest PWV\,---\,which is crucial for THz observations \citep{2016NatAs...1E...1S}, long dark nights \citep{2010AJ....140..602Z}, a high fraction of clear nighttime \citep{2021MNRAS.501.3614Y}, a shallow boundary layer \citep{2008SPIE.7014E..6IB,2010PASP..122.1122B}, strong temperature inversion and stable atmosphere \citep{2014PASP..126..868H, 2019PASP..131a5001H}, and superb seeing for optical and infrared observations \citep{2020Natur.583..771M}.

In the near-infrared, \cite{2023MNRAS.521.5624Z} reported that the sky at Dome A is as dark as South Pole in the $J$, $H$, and $K_s$ bands based on preliminary measurements from several nights in April 2019. This is not surprising given the study at the South Pole by \cite{1996PASP..108..721A} and the fact that Dome A has a higher altitude and lower temperatures than the South Pole and that thermal IR emission, which is significant in $K_s$ band, is greatly reduced by the low ambient temperature. 
The extremely low PWV at Dome A reduces absorption around 1.4\,\textmu m and 1.9\,\textmu m, opening new ground-based windows for observation \citep{2012PASP..124...74S}. The Kunlun Infrared Sky Survey \citep[KISS,][]{2016PASA...33...47B} was proposed to explore the dynamic universe in $K_{\mathrm{dark}}$ band, but was unable to proceed due to export restrictions on the detector. \citet{2024PASP..136k5002L} developed a 15\,cm near-infrared telescope with a FoV of 0.87\degr $\times$ 0.69\degr\,using an IMX990 InGaAs camera; this was installed at Dome A in January 2024 and conducted daytime observations, achieving a detection limit of $J = 10$\,mag with an effective exposure time of 175\,s \citep{2025AJ....169..228Y}.

Herein, we present the Antarctic Infrared Binocular Telescope (AIRBT), a time-domain survey pathfinder for Dome~A. The AIRBT consists of two 15~cm optical tube assemblies (OTAs) with commercial InGaAs cameras, providing a FoV of 1.22\degr\,$\times$\,0.97\degr. By simultaneously imaging in the $J$ and $H$ bands, AIRBT aims to study bright variables, monitor the sky background at Dome A over the long term, and lay the technological groundwork for future large telescopes. AIRBT was installed at Dome A in January 2023 by the 39th Chinese National Antarctic Research Expedition (CHINARE 39). Due to focus issues in the first year, the data reduction requires further processing. After commissioning, AIRBT began survey observations in $J$ and $H$ bands with good operational status in 2024.

In this paper, we describe the design of AIRBT and present test results from both laboratory and on-sky observations. The paper is organized as follows: Sect.~\ref{sec:equipment} provides an overview of the system, including the telescopes and InGaAs cameras. Sect.~\ref{sec:Test of cameras} presents the laboratory characterization of the detectors, while Sect.~\ref{sec:observation test} describes on-sky performance prior to deployment to Antarctica. We summarize the paper in Sect.~\ref{sec:Summary}.

\section{System overview}
\label{sec:equipment}

\begin{figure}
  \centering
  \includegraphics[height=5cm]{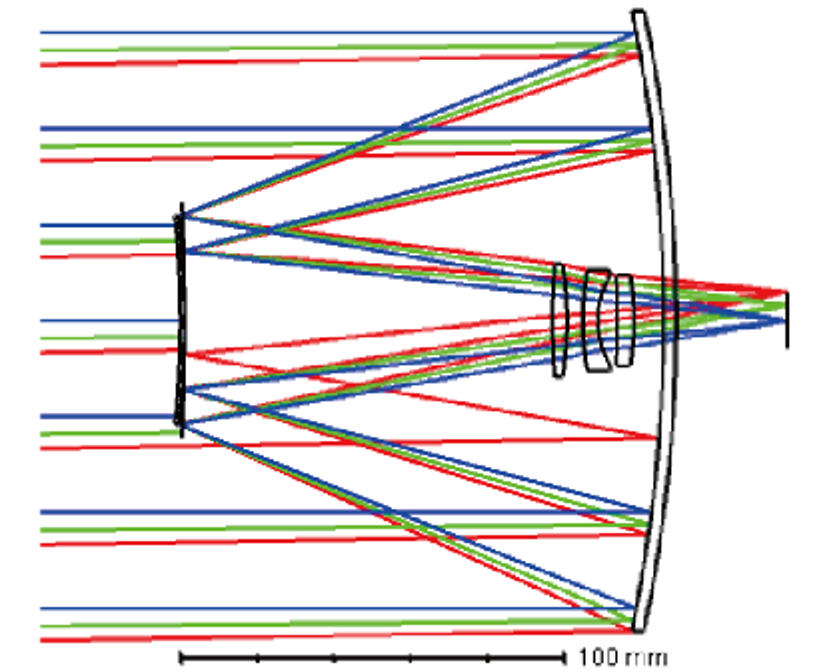}
  \caption{Optical design of the telescope, which is an RC design with lens correctors. The entrance pupil has a diameter of 15 cm and the overall focal ratio is $f/$3.}
  \label{fig:optical_path}
\end{figure}

AIRBT is optimized for wavelengths of 1.0\,--\,1.7\,\textmu m and the Antarctic environment. Each of the two OTAs adopts a Ritchey-Chrétien (RC) design with lens correctors, as shown in Fig.~\ref{fig:optical_path}. 
The OTAs have a diameter of 15\,cm and a focal ratio of $f$/3. The optical design achieves 80\% encircled energy within a radius of 15\,\textmu m across the full FoV, which has a diameter of 15\,mm (1.9\degr). The image quality remains stable at ambient temperatures from \textminus20\,\textcelsius\ to \textminus80\,\textcelsius, since the telescope structure is made of INVAR, a low-expansion material. Thus, the telescope requires only one focusing process during installation in the Antarctic summer when the ambient temperature is around \textminus20\,\textcelsius, and it will maintain sharp images throughout the year, even when the temperature drops to \textminus80\,\textcelsius. This eliminates the requirement for further focusing during the unmanned polar night. Electrically conductive Indium-Tin-Oxide (ITO) films are coated on the windows of AIRBT to prevent frost accumulation. The films are powered during observations to heat the windows to several degrees above the ambient temperature, thereby allowing frost to sublime. 

AIRBT uses two LD-SW640171550-C2-G cameras from Xi'an Leading Optoelectronic Technology Co., Ltd.
Each camera has a FPA0640P15F-17-T2 InGaAs focal plane array (FPA)\footnote{\url{https://www.clpt.com.tw/640x512P15/17603/}} with 640\,$\times$\,512 pixels. The pixel size is 15\,\textmu m, corresponding to 6.9\,arcsec\ on the sky, giving an effective FoV of 1.22\degr\,$\times$\,0.97\degr. These cameras utilize TEC to cool the FPA temperature to 30\,--\,40\,\textcelsius\ below the camera shell temperature. When operating at Dome A the FPA can readily achieve its optimal operating temperature of \textminus55\,\textcelsius.

The two cameras are equipped with $J$ and $H$ filters, respectively.
The spectral response of the InGaAs detectors is 0.9\,--\,1.7\,\textmu m with a quantum efficiency of $\geq$\,70\% at room temperature.
After cooling, the long wavelength cut-off becomes shorter, resulting in a response range of approximately 0.9\,--\,1.63\,\textmu m at \textminus40\,\textcelsius.
Therefore, the effective $H$ filter of AIRBT covers only the blue half of the 2MASS $H$ filter (1.51\,--\,1.79\,\textmu m) at Dome A. The wavelength range of the $J$ band filter of AIRBT is similar to that of the 2MASS $J$ filter (1.11\,--\,1.39\,\textmu m).

All hardware is designed for Dome A operating conditions. We choose these cameras not only for their performance but also for their suitability for Antarctic conditions. The cameras can operate normally down to \textminus40\,\textcelsius\, and even colder, accounting for their own heat generated during operation. Nevertheless, extreme temperatures during the Antarctic winter still pose operational risks. Consequently, we add thermal insulation and heaters for the cameras. The heaters are automatically controlled according to the camera temperature. The active thermal control ensures optimal operating temperature when the air temperature ranges from \textminus20\,\textcelsius\ to \textminus80\,\textcelsius, as verified by laboratory cold tests. 
All power and data transfer are provided through the PLATeau Observatory \citep[PLATO,][]{2009RScI...80f4501L, 2010EAS....40...79A} platform. A network cable, long enough to cover the 30-meter distance between AIRBT and its computer, also provides redundancy by enabling every computer on the local area network to communicate with the cameras. 

\section{Laboratory tests of cameras}
\label{sec:Test of cameras}

We characterized the cameras' performance through laboratory tests, including bias, readout noise, dark current, non-linearity, photon transfer curve (PTC), and gain. We followed the standard procedure for CCD tests and found that InGaAs cameras exhibit some differences from CCDs:
(1) the bias takes a short time (10\,\textmu s) to reach a stable level;
(2) the bias decreases with decreasing FPA temperature;
(3) the noise increases with integration time during PTC measurements, affecting the gain calculation.

The cameras feature 14-bit outputs with maximum count values of 16,383 in analog-to-digital units (ADUs). The signal chains operates in three gain modes (high, middle, and low), each with distinct parameters while maintaining similar trends. In this section, we present detailed results from the $J$ band camera, as the performance of the $H$ band camera was comparable.

\subsection{Bias and readout noise}
\label{sec:bias}

For CCDs the bias is essentially constant for each pixel, and a bias frame is obtained from a dark frame with the shortest integration time, or at least one short enough to make the accumulated dark current negligible. The noise in these frames is also relatively constant, corresponding to the readout noise. Additionally, the bias is almost invariant with temperature. However, InGaAs detectors behave differently.

\begin{figure}
  \centering
  \includegraphics[height=15cm]{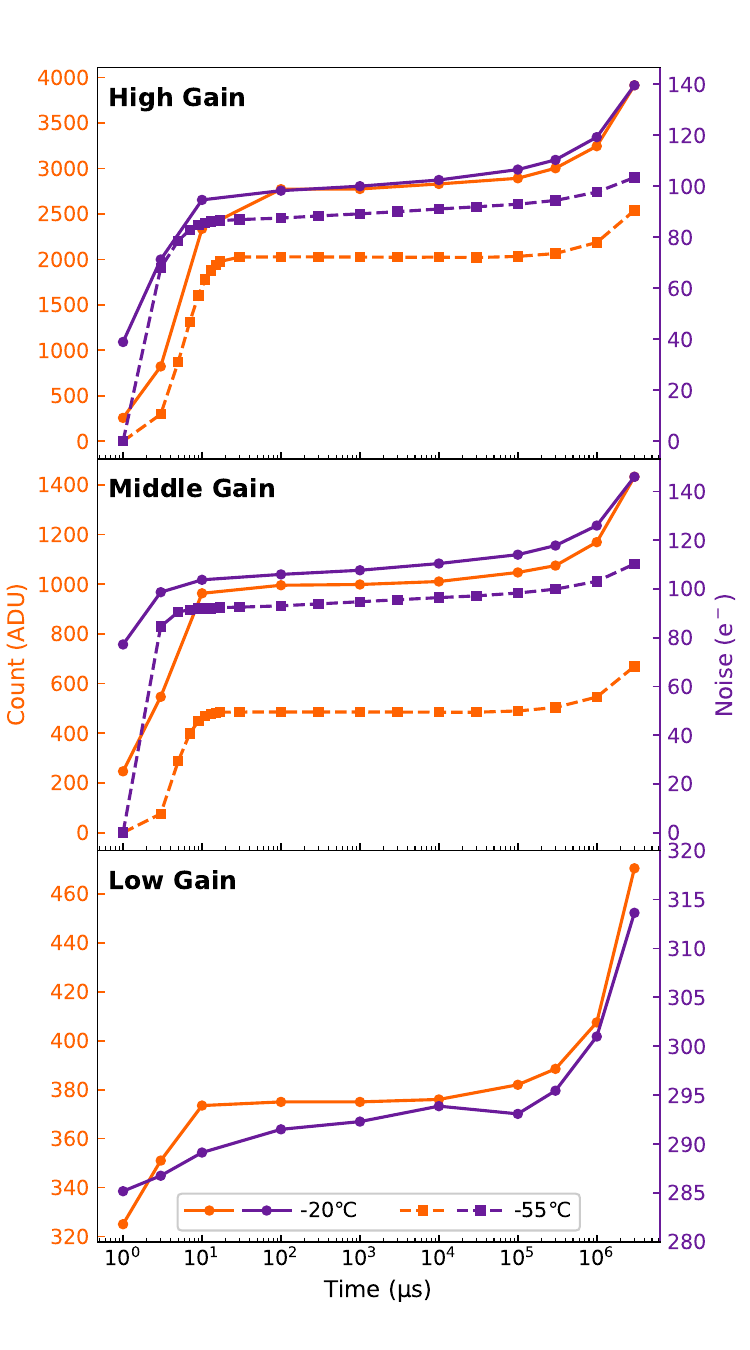} 
  \caption{Counts (orange) and noise (purple) versus time from dark frames. Solid curves represent results with the FPA temperature around \textminus20\,\textcelsius, while dashed curves represent FPA temperature around \textminus55\,\textcelsius. We show that both the bias and readout noise take time to stabilize and decrease with lower FPA temperature.
  }
  \label{fig:st}
\end{figure}

Firstly, both the bias and readout noise require a sufficiently long integration time to reach a stable state, as illustrated in Fig.~\ref{fig:st}. The black solid lines represent median counts (bias + signal) in ADUs from short dark frames taken with the camera at room temperature and the FPA cooled to \textminus20\,\textcelsius, typical for most sites. For the shortest integration time (1\,\textmu s) at middle and high gains, many pixels have zero counts. We suspect this is due to the frequency response of the signal chain for these gains. As the integration time increases from 1\,\textmu s to about 10\,\textmu s in middle gain mode, the median count increases rapidly from 200 ADU to 999 ADU. Similarly, the median noise (blue solid lines) rises from 0 to 107\,$e^{-}$. After this, the count stabilizes, while the noise continues to increase slowly as the accumulated dark current becomes significant. This indicates that the bias of InGaAs detectors takes time to stabilize, so it is important to select an appropriate integration time for deriving a bias frame. For our cameras, we chose 1\,ms dark frames as the bias. The bias frame exhibits striped patterns, as shown in Fig.~\ref{fig:bias_and_dark}. This temporally stable pattern can be removed from science images using standard bias subtraction.

Secondly, both the bias and readout noise decrease with the FPA temperature. The dashed lines in Fig.~\ref{fig:st} show counts and noises with the camera at low temperature, with the FPA cooled to \textminus55\,\textcelsius, typical for Antarctica. In this case, the bias level decreases from 2771 (999) to 2025 (485) ADU in high (middle) gain mode. The effect is even more pronounced in low gain mode, where the bias drops from 375 to 73 ADU, and it takes significantly longer (several seconds) to stabilize. For readout noise, the effective values in high, middle, and low gain modes are 96 (83), 107 (93), and 290 (253)\,$e^{-}$ at \textminus20 (\textminus55)\,\textcelsius\ , respectively.

\subsection{Dark current}
\label{sec:dark} 

\begin{figure*}
  \centering
  \includegraphics[height=5cm]{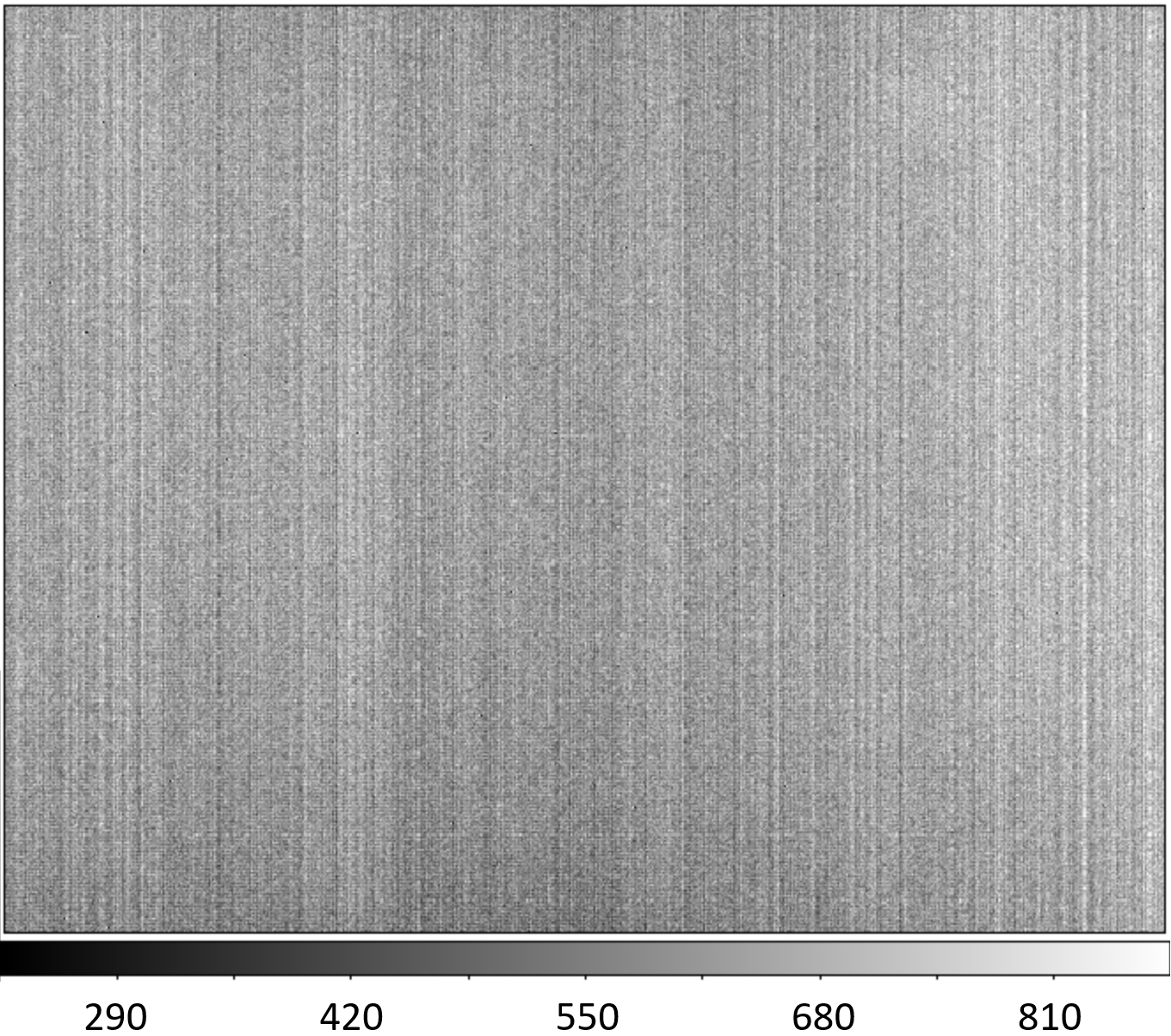}\quad
  \includegraphics[height=5cm]{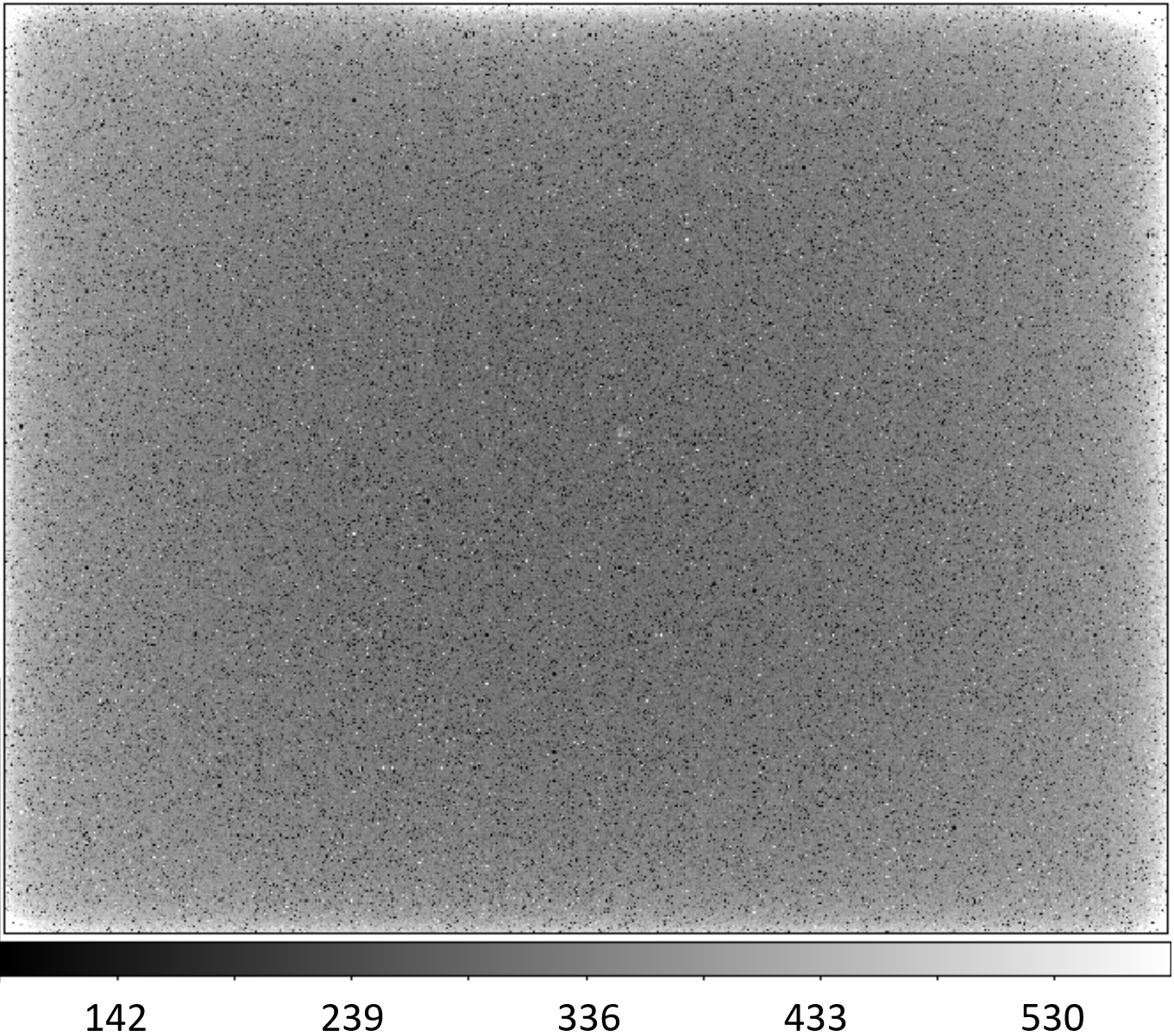}\\
  \includegraphics[height=5cm]{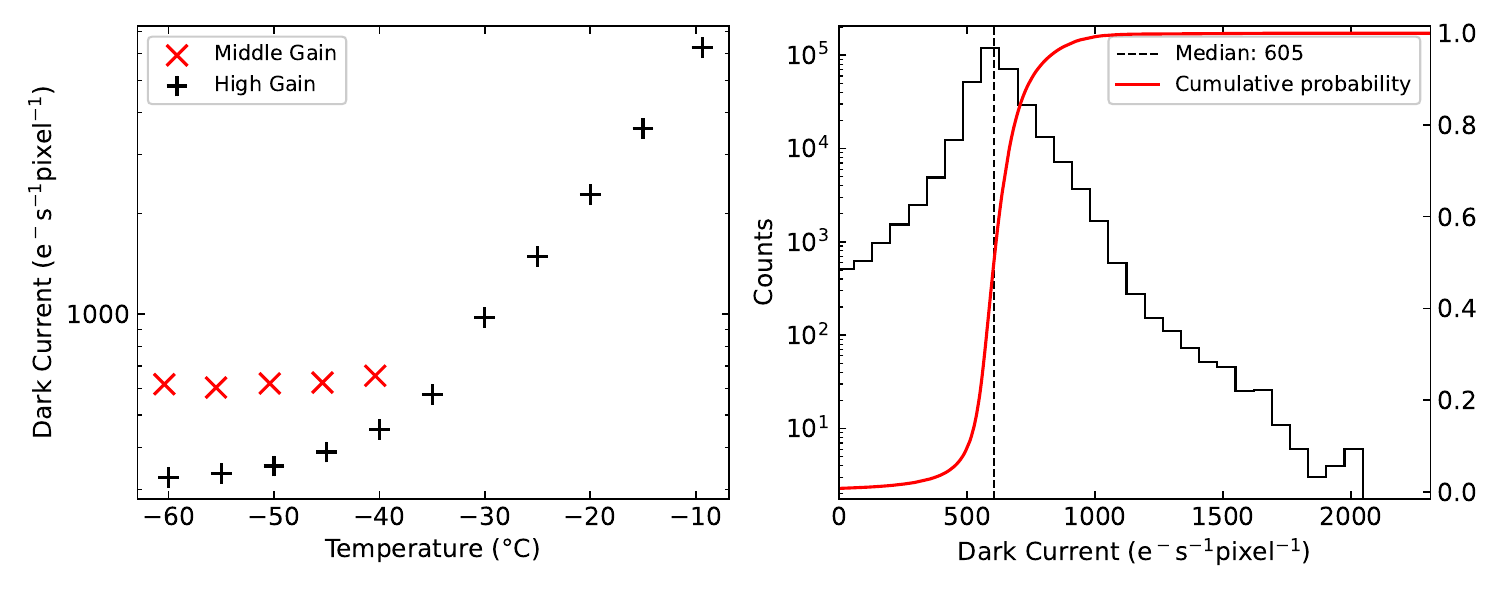}\quad
  \caption{Bias and dark current. Upper left: Bias frame with temporally stable stripe patterns. Upper right: Dark current frame with a 5\,s integration at \textminus55\,\textcelsius\ in middle gain mode. Lower left: Dark current versus temperature in high gain (black) and middle gain (red) modes. The dark current decreases exponentially with temperature until \textminus40\,\textcelsius\ . The dark current in middle gain mode is higher than that in high gain mode. Lower right: Distribution and cumulative curve of middle gain dark current at \textminus55\,\textcelsius, following approximate normal statistics. All data are acquired in middle gain mode.
  }
  \label{fig:bias_and_dark}
\end{figure*}

To deeply cool the FPA with its TEC, we put the camera in a low-temperature test chamber with a temperature of \textminus40\,\textcelsius. 
The lower left panel of Fig.~\ref{fig:bias_and_dark} displays the median dark current temperature dependence of the high and middle gain mode. The dark current decreases sharply with temperature in high gain mode, then slows beyond \textminus40\,\textcelsius\ . A similar trend is observed in middle gain mode, though not depicted here. The dark current is higher in middle gain compared to high gain mode. The stable operating temperature for middle gain mode is \textminus55\,\textcelsius\ , with a dark current of 605\,$e^{-}$\,s$^{-1}$\,pixel$^{-1}$ in middle gain mode.

Fig.~\ref{fig:bias_and_dark} presents a 5\,s dark frame at \textminus55\,\textcelsius\ in middle gain mode. The pixels near the edge show 50\% higher dark current than those of normal pixels. The histogram of dark current follows approximately a normal distribution, with 90\% below 750\,$e^{-}$\,s$^{-1}$\,pixel$^{-1}$. 

\subsection{Nonlinearity}
\label{sec:linearity}

\begin{figure*}
    \centering
    \includegraphics[height=4.6cm]{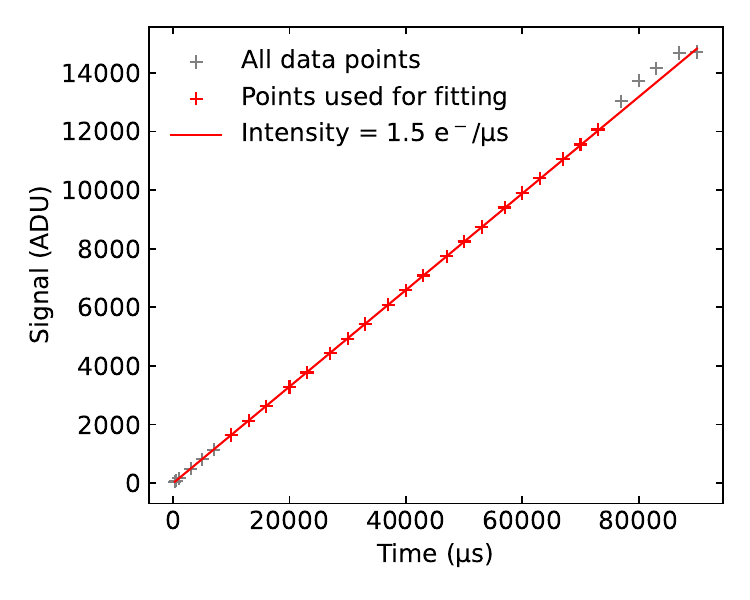}\quad
    \includegraphics[height=4.6cm]{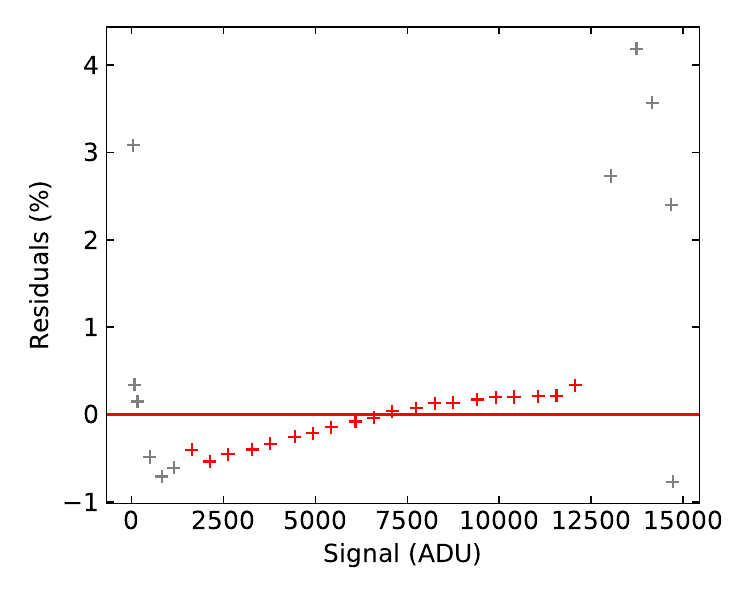}\quad
    \includegraphics[height=4.6cm]{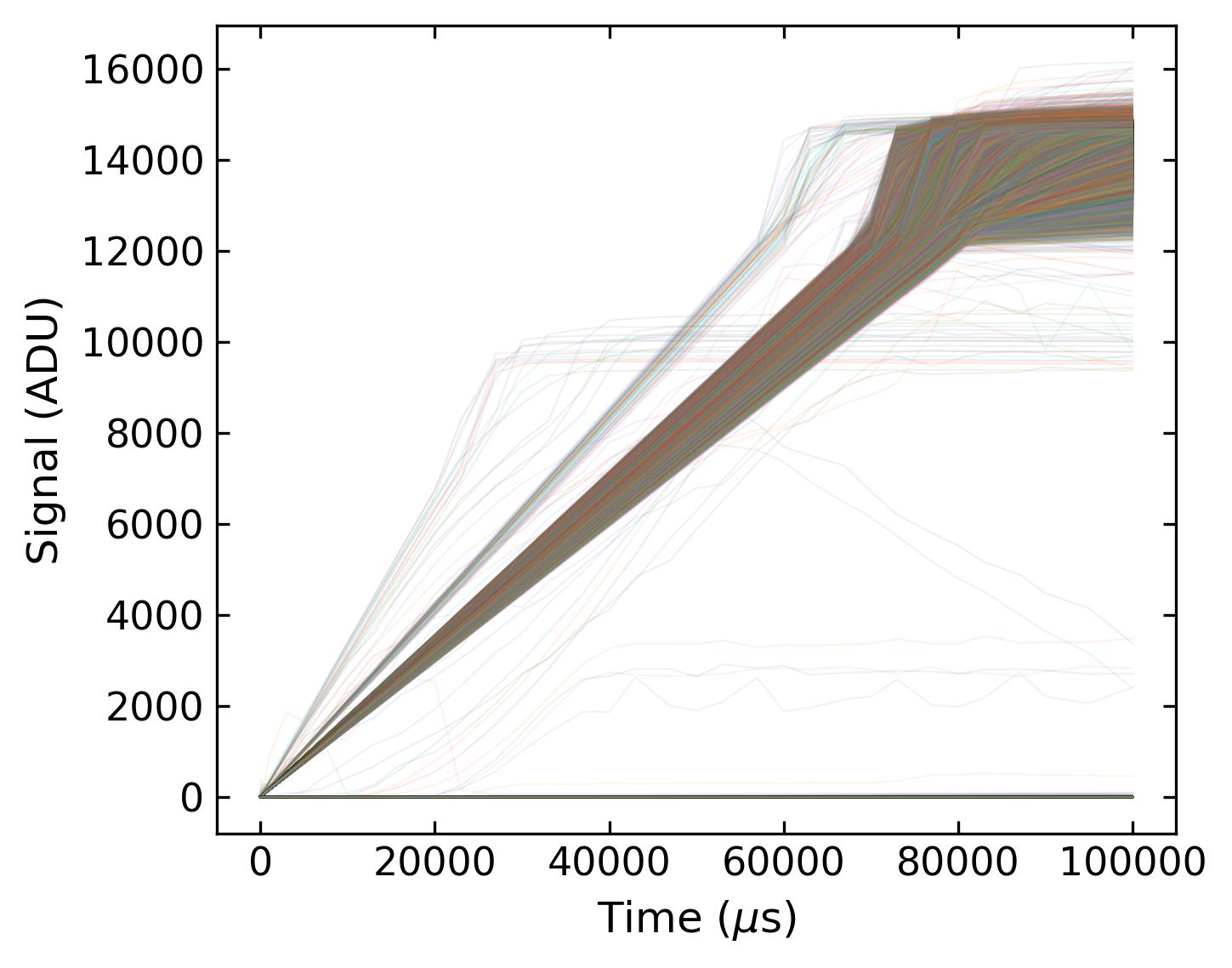}
    \caption{Left: Overall linearity. Middle: Nonlinearity, which stays below 1\% between 1000 and 12,500 ADU. Right: Single-pixel linearity, where 117 pixels show no response or nonlinear behavior. All data were acquired in middle gain mode.
    }
    \label{fig:159_linearity_profile}
\end{figure*}

To investigate nonlinearity, we took flat-field frames with increasing exposure times in middle gain mode and plot the mean signal versus time in Fig.~\ref{fig:159_linearity_profile}. The plot shows high linearity between 1000 and 12,500 ADU, with nonlinearity less than 1\%. However, 117 pixels (0.04\% of the total) exhibit higher nonlinearity, smaller linear ranges, or no response to light.

\subsection{Photon transfer curve and gain}
\label{sec:photon transfer curve}

The PTC is the relationship between variance and signal from flat-field frames. The variance consists of two noises: the readout noise $\sigma^2_{\mathrm{rdn}}$, which is relatively constant, and the photon shot noise, which follows a Poisson distribution. If the variance and signal are measured in units of ADU, the variance is given by:
\begin{equation}
\mathrm{Variance} = \sigma^2_{\mathrm{rdn}} + \frac{\mathrm{Signal}}{\mathrm{Gain}}.
\end{equation}
Therefore, the relationship is linear, and the slope is typically used to derive the gain. We apply the standard method to calculate variance by subtracting consecutive frames to eliminate the impact of fixed pattern noise.

\begin{figure*}
  \centering
  \includegraphics[height=5cm]{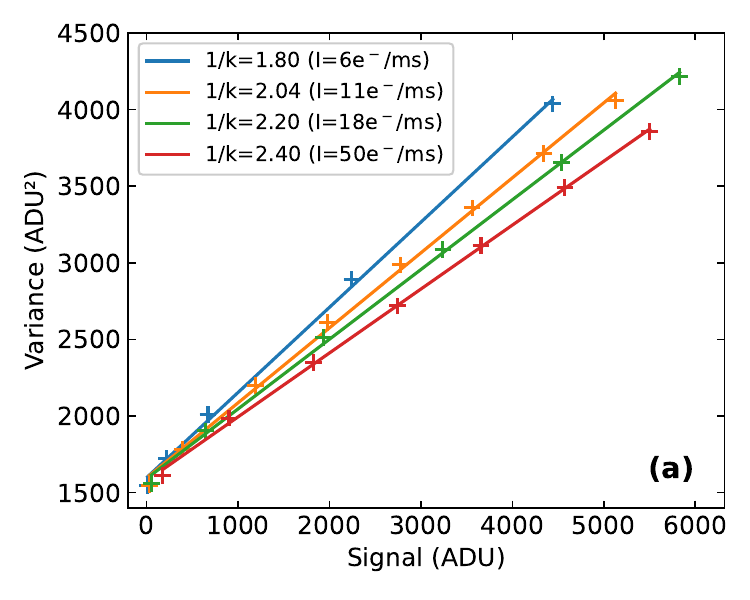}\quad
  \includegraphics[height=5cm]{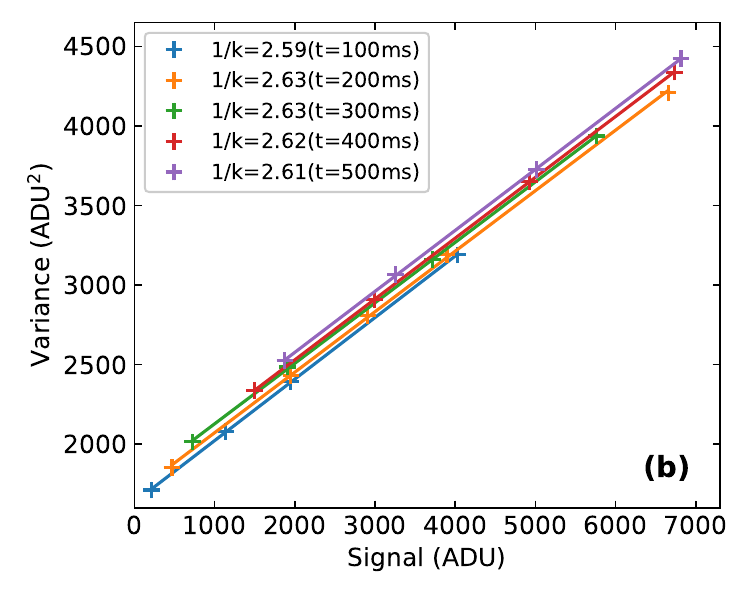}\\
  \includegraphics[height=5cm]{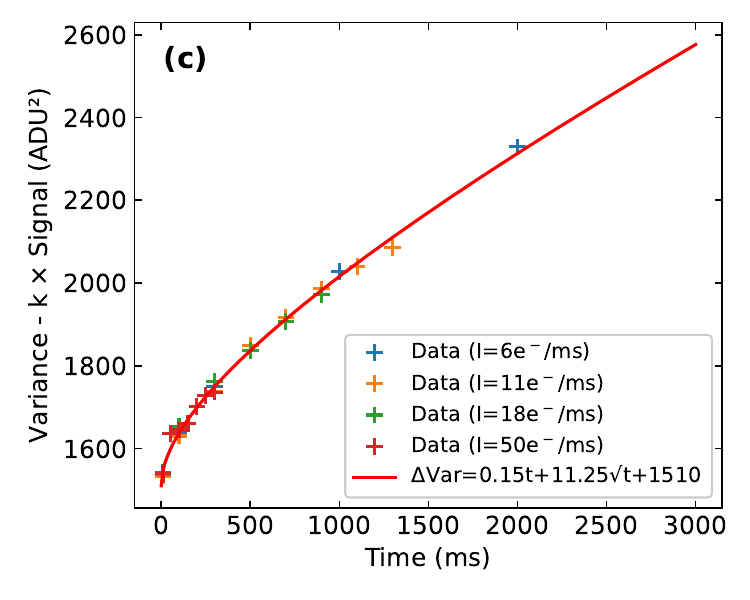}\quad
  \includegraphics[height=5cm]{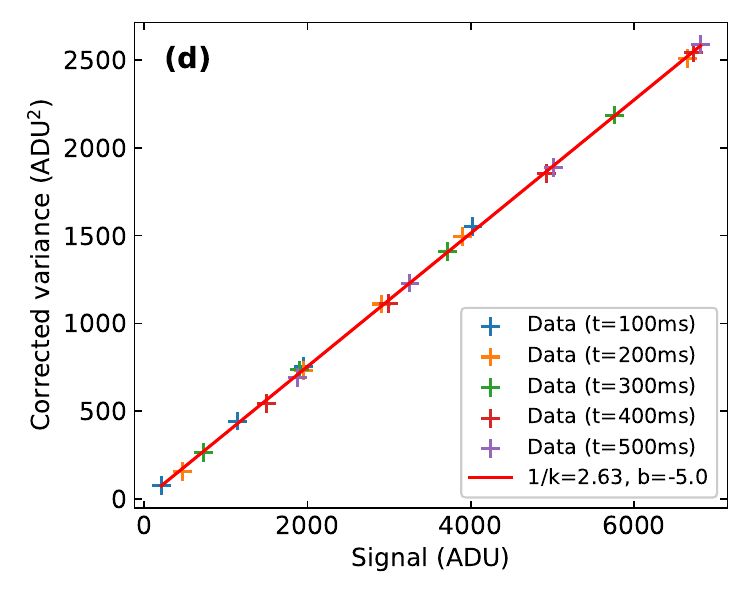}
  \caption{The PTCs and gain estimation. 
  (a) PTCs obtained by increasing exposure time with a stable light source. Different curves represent varying intensity levels. Variance for identical counts increases with lower light intensity, requiring longer integration time and resulting in higher variance. 
  (b) PTCs for fixed exposure time with increasing light intensity. Different curves correspond to different exposure times. All PTC slopes remain consistent, indicating a gain of 2.6\,e$^-$/ADU. Longer integration times lead to greater interference, suggesting that additional noise accumulates over time.
  (c) Extra noise obtained by subtracting photon shot noise from the PTCs in (a).
  (d) Noise from the PTCs in (b) with extra noise subtracted. All data points fall on the same line.
  }
  \label{fig:PTC}
\end{figure*}

However, we find extra noise in the PTCs in all three gain modes. Limited by the highest intensity of our light sources, we used high gain mode as an example. We obtain PTCs in the traditional way by increasing the exposure time with a stable light source to increase the signal. Fig.~\ref{fig:PTC}(a) shows a set of PTCs, each obtained with a different light brightness. The PTCs are linear but surprisingly have varying slopes. The brighter the light, the gentler the slope, resulting in inconsistent gain results. For the same signal, the variance is larger for fainter light, i.e., longer exposure times. Therefore, we hypothesize the presence of extra noise that increases with integration time.

To confirm this, we alter the method for obtaining PTCs. Since the extra noise is time-dependent, we fix the exposure time and increase the signal by brightening the illumination to obtain a PTC. We obtain another set of PTCs for different fixed exposure times. The results shown in Fig.~\ref{fig:PTC}(b) confirm our hypothesis. The PTCs exhibit consistent slopes, or equivalently, consistent gain values, because each point in the same PTC has constant extra noise. On the other hand, the intercepts of the PTCs increase with integration time, proving that the noise increases with integration time.

To derive this extra noise, shot noises from both photoelectrons and dark current\,---\,calculated using the gain values from the second set of PTCs\,---\,were subtracted from the first set of PTCs. The resultant remaining noise is plotted in Fig.~\ref{fig:PTC}(c). The time dependence of the noise can be well fitted by
\begin{equation}
    \Delta \mathrm{Var} = 0.15 t + 11.25 \sqrt{t} + 1510,
    \label{eq:noise}
\end{equation}
where ${\Delta \mathrm{Var}}$ corresponds to the additional noise plus readout noise, and $t$ represents the integration time in units of ms. The constant term is 1510 ADU$^2$, i.e., a readout noise of 101\,e$^-$, which agrees with the value derived from bias in Sect.~\ref{sec:bias}.

To verify Eq.~\ref{eq:noise}, we subtract the time-dependent noise from the second set of PTCs. The corrected variance versus signal is shown in Fig.~\ref{fig:PTC}(d). All data points fall on the same line, with a slope identical to the original. The assumption of extra noise explains both sets of PTCs well, though its origin remains unclear. The time dependence consists of two terms: a linear term and a square-root term, implying multiple possible causes. \cite{2017InPhT..85...74Y} describes noise components of InGaAs detectors, including time-dependent shot noise, detector thermal noise, 1/f noise, amplifier thermal noise, reset noise from the sample circuit, and fixed pattern noise et al. The additional noise may represent a combination of several of these components. However, their analysis does not account for a noise contribution whose variance follows a half-order dependence on time, a phenomenon that requires further study.

To accurately measure the camera's gain from PTCs, we recommend using fixed exposure times and variable light intensity. In this way, we measured gain values for all gain modes, yielding values of 2.6, 8.9, and 79\,$e^-$/ADU for the high, middle, and low gain modes, respectively. One should also take into account the extra noise to esimate the signal-to-noise ratio (S/N) in photometry. 

\subsection{Summary of camera tests}
\label{sec:summary_camtest}

We summarize the laboratory test results for the three gain modes in Table~\ref{tab:detector}. Accordingly, we adopt the middle gain mode for its balance between readout noise and saturation capacity. We found that stacking 9 middle-gain frames (each with exposure time $t$) achieves a dynamic range and readout noise level equivalent to a single low-gain frame with an exposure time of 9$t$. The multi-frame strategy provides higher temporal resolution and can remove contamination from cosmic rays or satellite trails. However, this is not the case for the high gain mode due to its high effective readout noise. Therefore, the middle gain mode is optimal. Under the working conditions at Dome A, the middle gain mode has a readout noise of 93\,$e^-$, dark current of 605\,$e^{-}$\,s$^{-1}$\,pixel$^{-1}$, and saturation capacity of 146 ke$^-$. The nonlinearity is better than 1\%, with a fraction of bad pixels of 0.04\%.

\begin{table*}
  \centering
  \caption{The InGaAs detector specifications of high, middle, and low gain modes at \textminus55\,\textcelsius, operating temperature.}
  \begin{tabular}{c|ccccc}
    \hline
    \shortstack{Mode \\ \, \\ \, } & \shortstack{Gain \\ (e$^-$/ADU)} & \shortstack{Bias \\ (ADU)} & \shortstack{Readout noise \\ (e$^-$)} & \shortstack{Dark current 
    \\ (e$^-$s$^{-1}$pix$^{-1}$)} & \shortstack{Saturation capacity \\ (ke$^-$)} \\
    \hline
    High & 2.6 & 2025 & 83 & 395 & 43 \\
    Middle & 8.9 & 485 & 93 & 605 & 146 \\
    Low & 79 & 73 & 253 & 545 & 1294 \\
    \hline
  \end{tabular}
  
  \label{tab:detector}
\end{table*}

Beyond these specifications, one should pay attention to the divergent behaviors between InGaAs cameras and CCDs. The bias of InGaAs cameras needs a short time to reach the stable level, and the bias decreases as the temperature decreases. Extra noise increases over time in the PTCs, and the slope will change if the signal increases with longer exposure time. An unbiased gain can be derived from a PTC by fixing the exposure time and increasing the light intensity to vary the signal. These behaviors may be common in InGaAs detectors, though to varying degrees, as we have observed them across multiple cameras from different manufacturers \citep[e.g.,][]{2024SPIE13103E..1MD}.

\section{On-sky performance tests}
\label{sec:observation test}

\begin{figure}
  \centering
  \includegraphics[height=8cm]{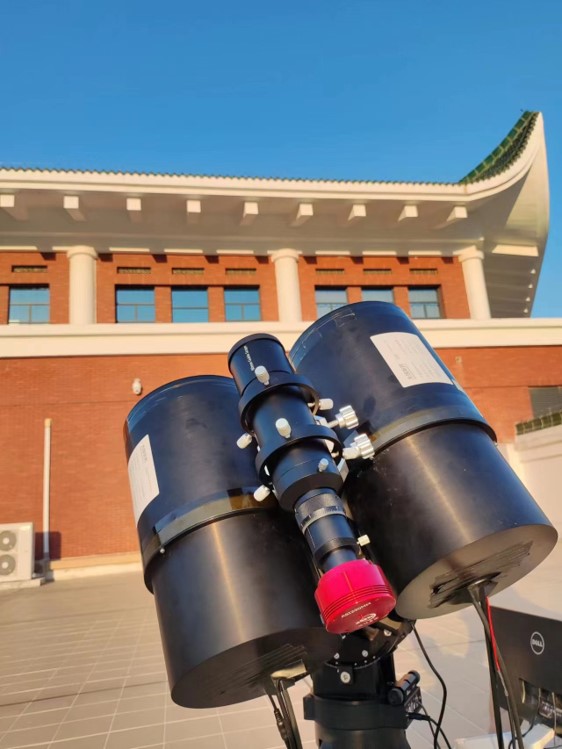}
  \caption{AIRBT at the Zhuhai campus of Sun Yat-sen University on October 11, 2022.}
  \label{fig:telescope}
\end{figure}

We performed on-sky tests of AIRBT at the Zhuhai campus of Sun Yat-sen University in October 2022 to verify the photometric performance of the whole system, as shown in Fig.~\ref{fig:telescope}. During observations, we obtained simultaneous $J$ and $H$ band images using the middle gain mode with 3\,s exposures, considering the bias level, dark current, sky background, and saturation capacity mentioned in Sect.~\ref{sec:Test of cameras}. 
Our observations targeted 20 fields in the Galactic plane to ensure enough stars in the images. The image depth was limited by the higher FPA temperature of \textminus 14.4\textdegree C, since the ambient temperature was above 20\textdegree C. This resulted in a dark current of 6,500\,$e^{-}$\,s$^{-1}$\,pixel$^{-1}$, about nine times higher than expected in Antarctica.

\subsection{Image pre-processing}

\begin{figure*}
  \centering
  \includegraphics[height=5cm]{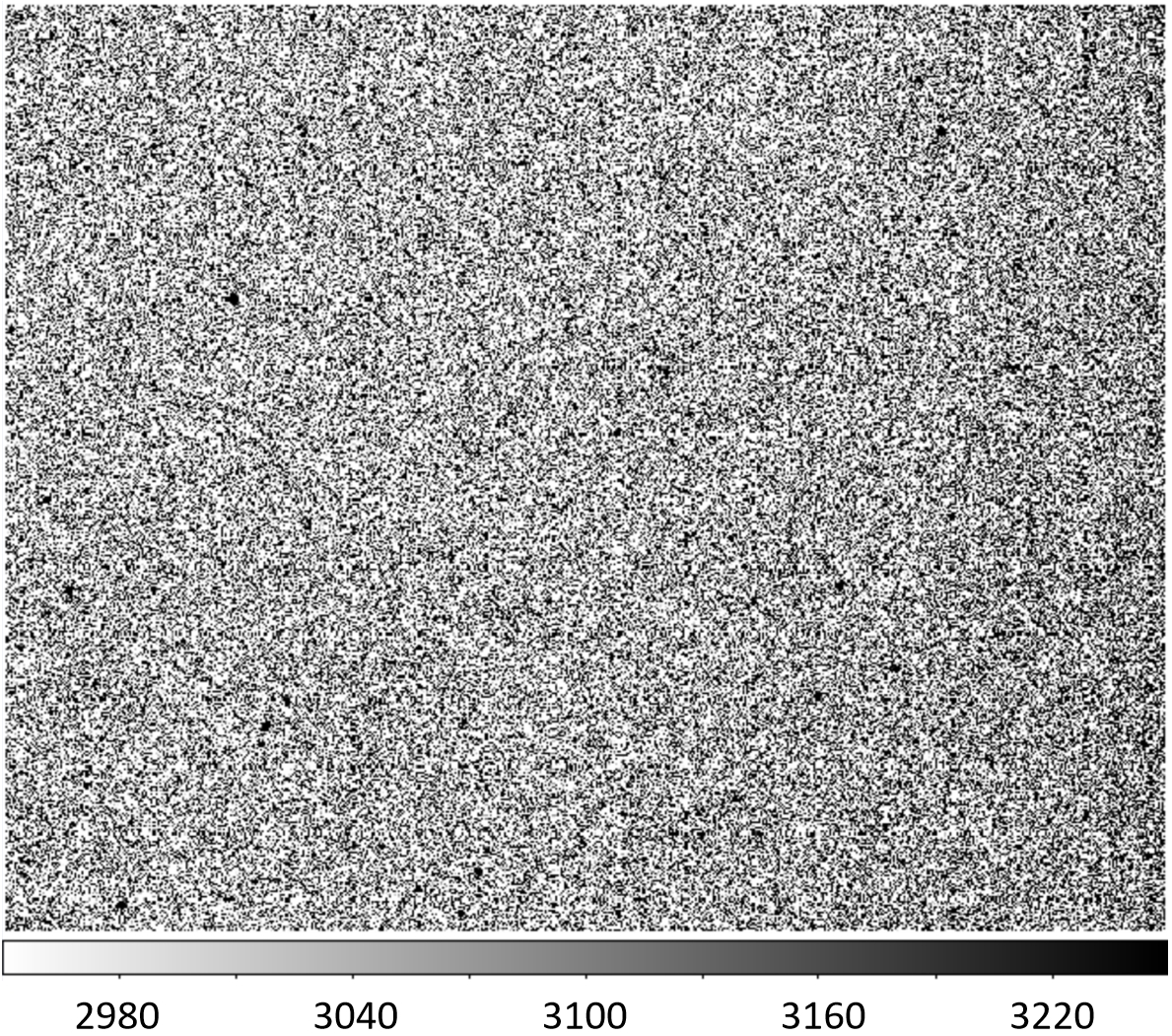}\quad
  \includegraphics[height=5cm]{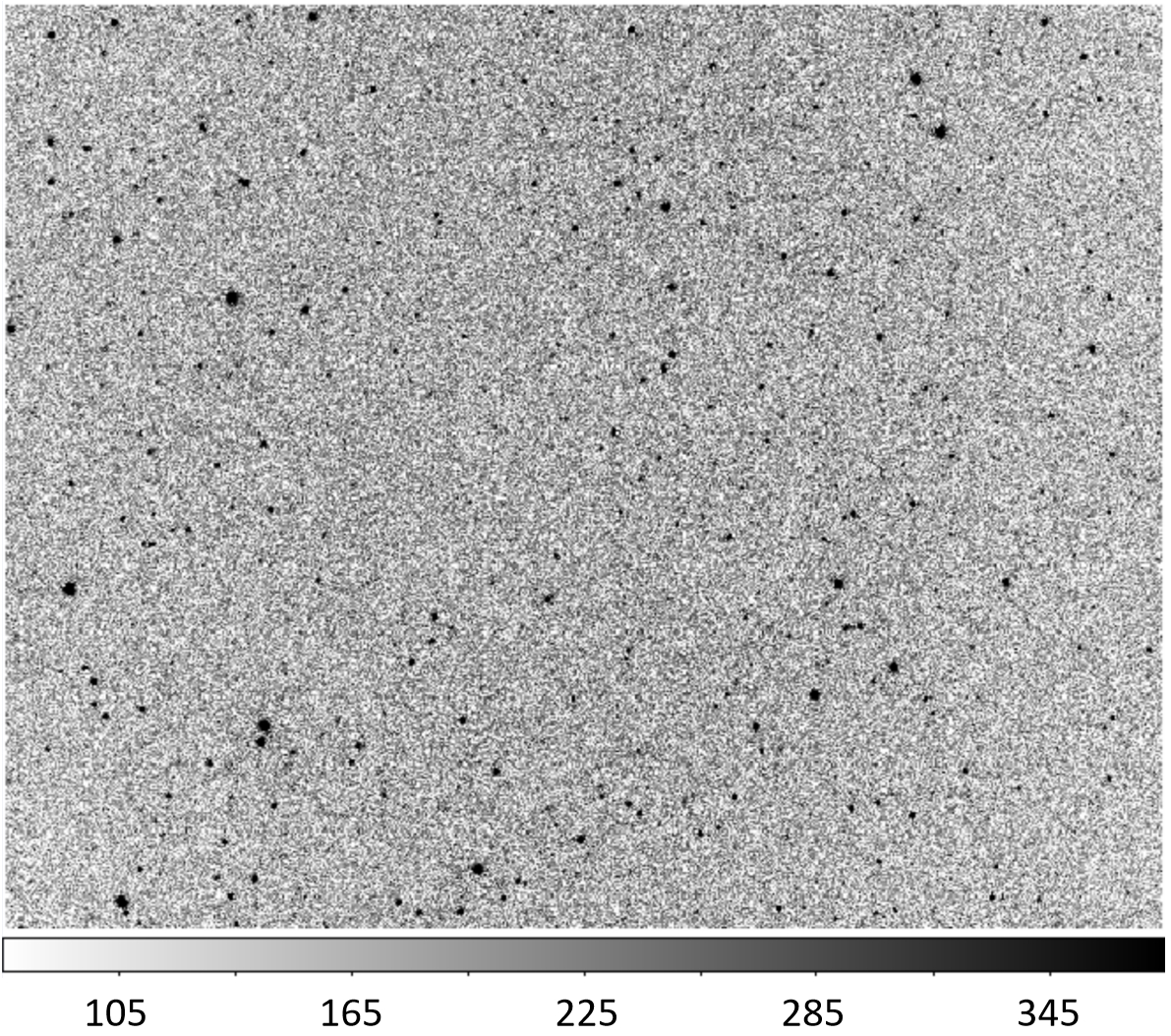}\quad
  \includegraphics[height=5cm]{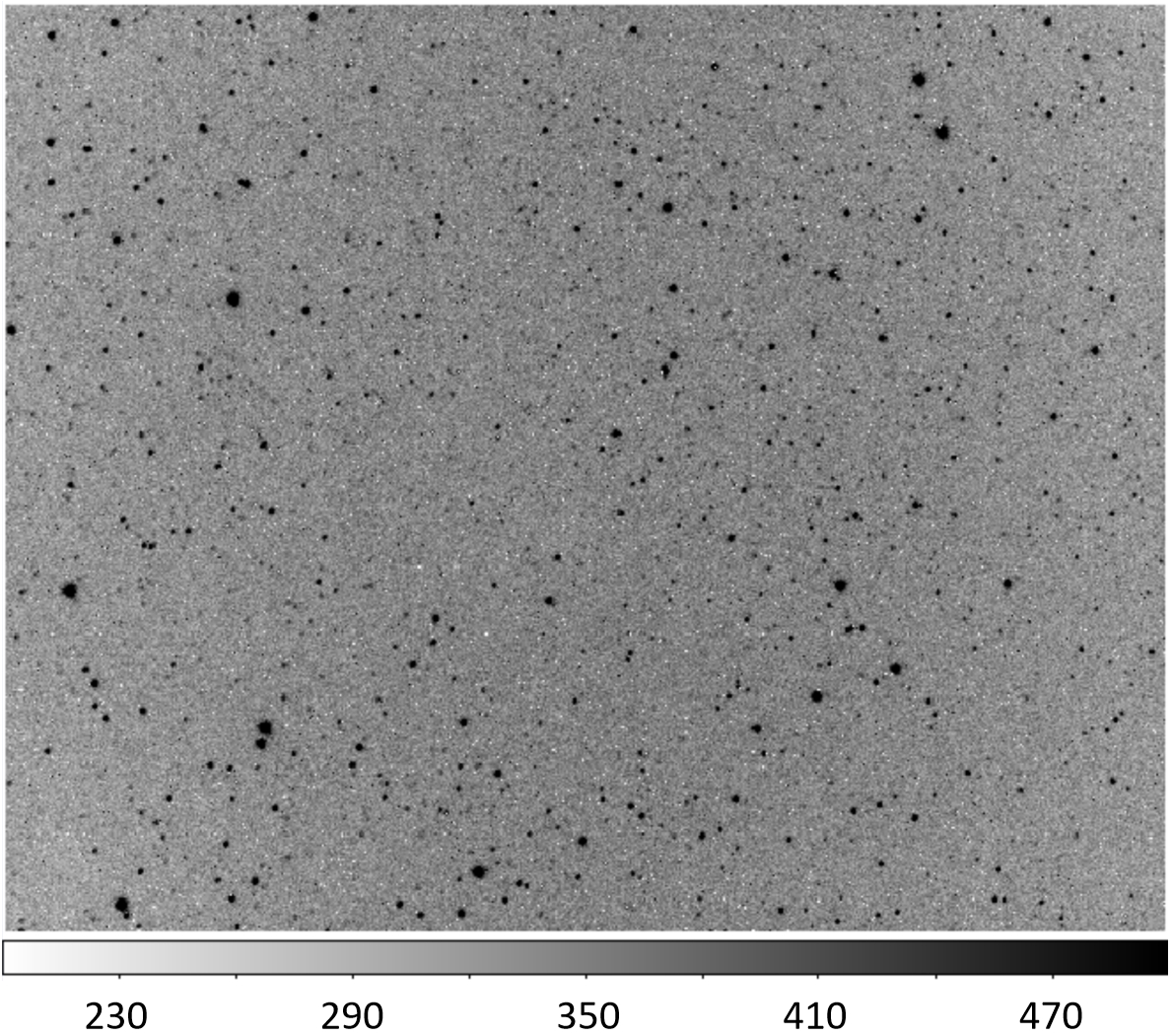}
  \caption{The effect of dark frame subtraction. Left: an example of raw frames from the observation. Middle: the image after subtracting the dark current captured in the laboratory. Right: the image after subtracting the dark current scaled by warm pixels. The background RMS decreases to 26 ADU, approaching the theoretical noise limit.
  }
  \label{fig:image}
\end{figure*}

We corrected the raw images following bias and dark current calibrations. However, obvious residuals of dark current patterns remained, as shown in Fig.~\ref{fig:image}. This indicates that the dark current levels fluctuated slightly, likely due to instability of the TEC cooling and changes in the ambient temperature. To quantify the dark current in our scientific images, we scaled the dark frames obtained from laboratory tests. This scaling used a linear correction coefficient determined by the temperature-dependent ratio of the intensity difference between warm and normal pixels, as described in \citet{2014SPIE.9154E..1TM, 2018MNRAS.479..111M}. The correction yields negligible residuals, as demonstrated in Fig.~\ref{fig:image}, where the background RMS is close to the theoretical noise limit. The full width at half maximum (FWHM) of the image is about 2.8 pixels.

\subsection{Photometry and astrometry}

We performed aperture photometry using Source Extractor \citep{1996A&AS..117..393B}. We assigned weights to all pixels. For bad pixels, we set the weight to zero to mask them. For normal pixels, we calculated the weight using:
\begin{equation}
    weight = \frac{1}{variance} = \frac{1}{\sigma^2_{\text{rdn}}  + \sigma^2_{\text{Poisson}}},
\end{equation}
where $\sigma^2_{\text{rdn}}$ denotes the readout noise variance from the bias, and $\sigma^2_{\text{Poisson}}$ indicates the Poisson noise variance from the dark current, sky background, and source flux. We used aperture diameters of 2, 4, and 6 pixels, corresponding to 13.8, 27.6, and 41.4 arcsec, respectively.

\begin{figure}
    \centering
    \includegraphics[height=6cm]{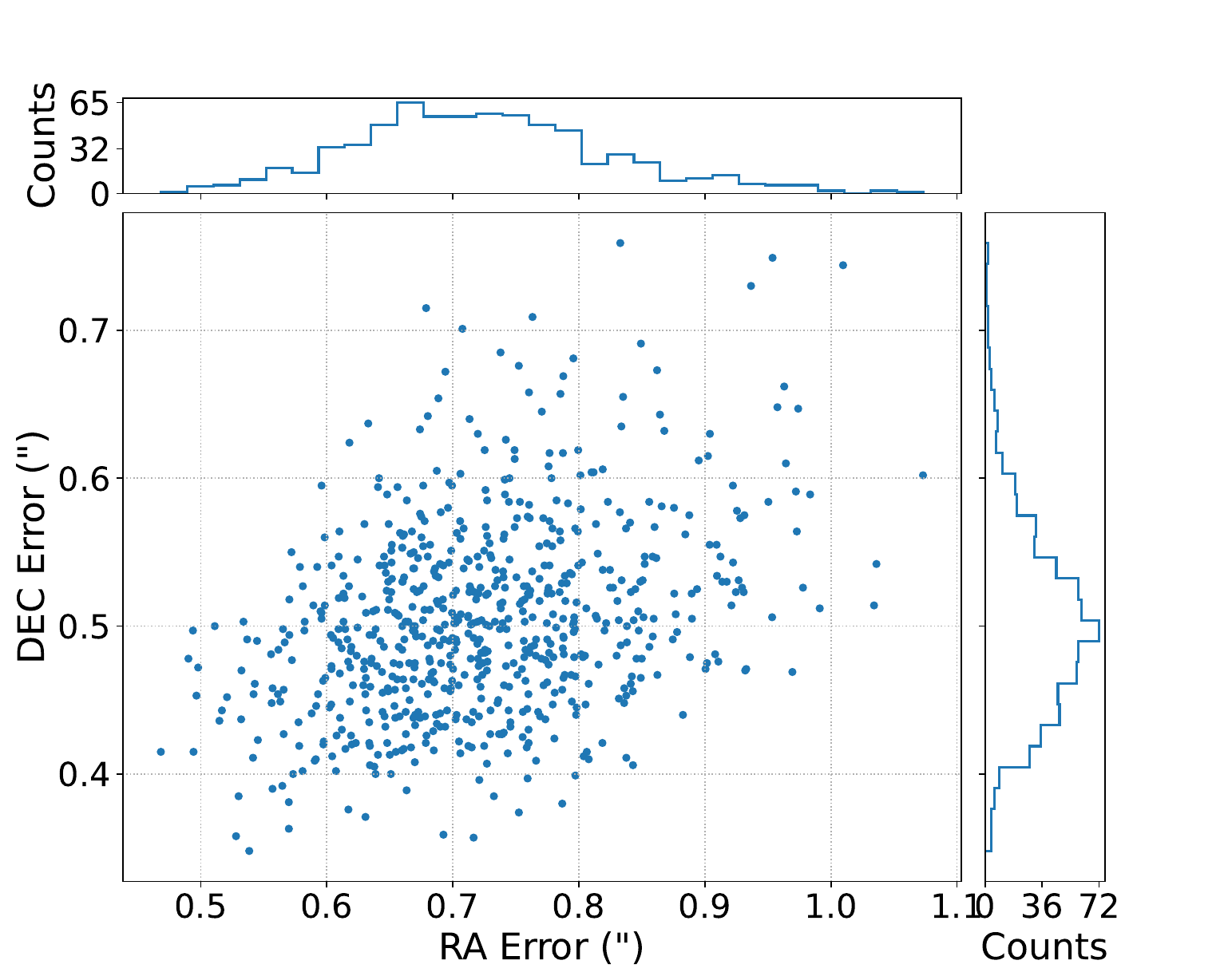}
    \caption{Astrometry precision measurements for 682 frames across all observation fields.}
    \label{fig:159-scamp-error}
\end{figure}

The World Coordinate System (WCS) was established using SCAMP \citep{2006ASPC..351..112B}, with astrometric calibration performed against the 2MASS catalog \citep{2006AJ....131.1163S}. Fig.~\ref{fig:159-scamp-error} shows median positional errors of 0.7 arcsec in right ascension (RA) and 0.5 arcsec in declination (DEC). Given the pixel scale of 6.9 arcsec per pixel, these uncertainties correspond to 0.1 pixel precision.

\subsection{Flux calibration}
\label{sec:Flux calibration}

We performed flux calibration using the formula:
\begin{equation}
    \text{Magnitude} = -2.5 \log(\text{Flux}) + \text{Zero-point},
\end{equation}
where \(\text{magnitude}\) is relative to the 2MASS catalog, and \(\text{flux}\) represents the total signal in ADU for a 3 second exposure frame within the aperture. We utilized data from all 20 fields and selected the frame with optimal image quality, resulting in 6,677 targets. The selection criteria included an FWHM of \(\leq 3\) pixels, sky background level (3\(\sigma\) clipping), dark current (3\(\sigma\) clipping), and atmospheric extinction (3\(\sigma\) clipping).

\begin{figure*}
    \centering
    \includegraphics[height=4.7cm]{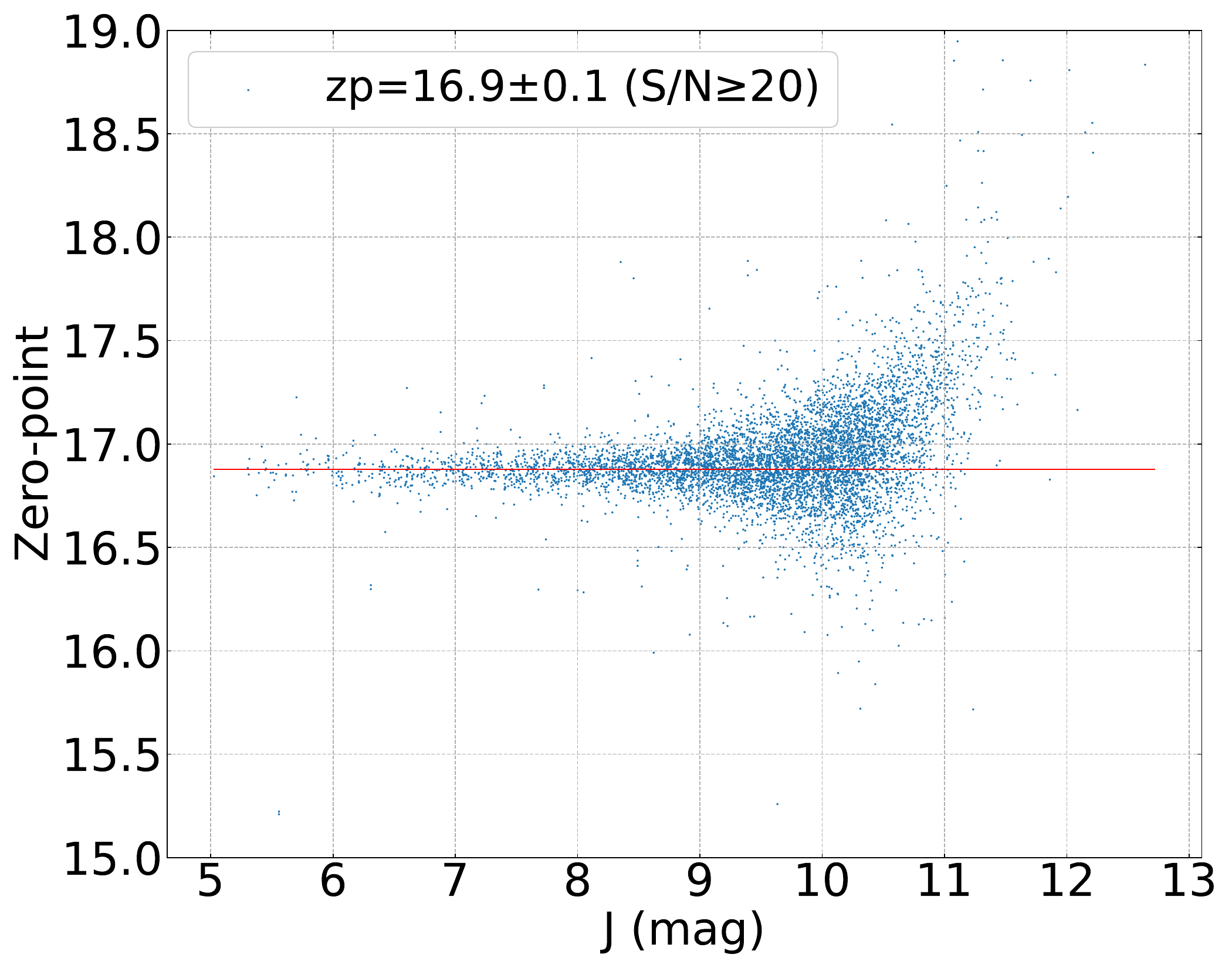}
    \includegraphics[height=4.7cm]{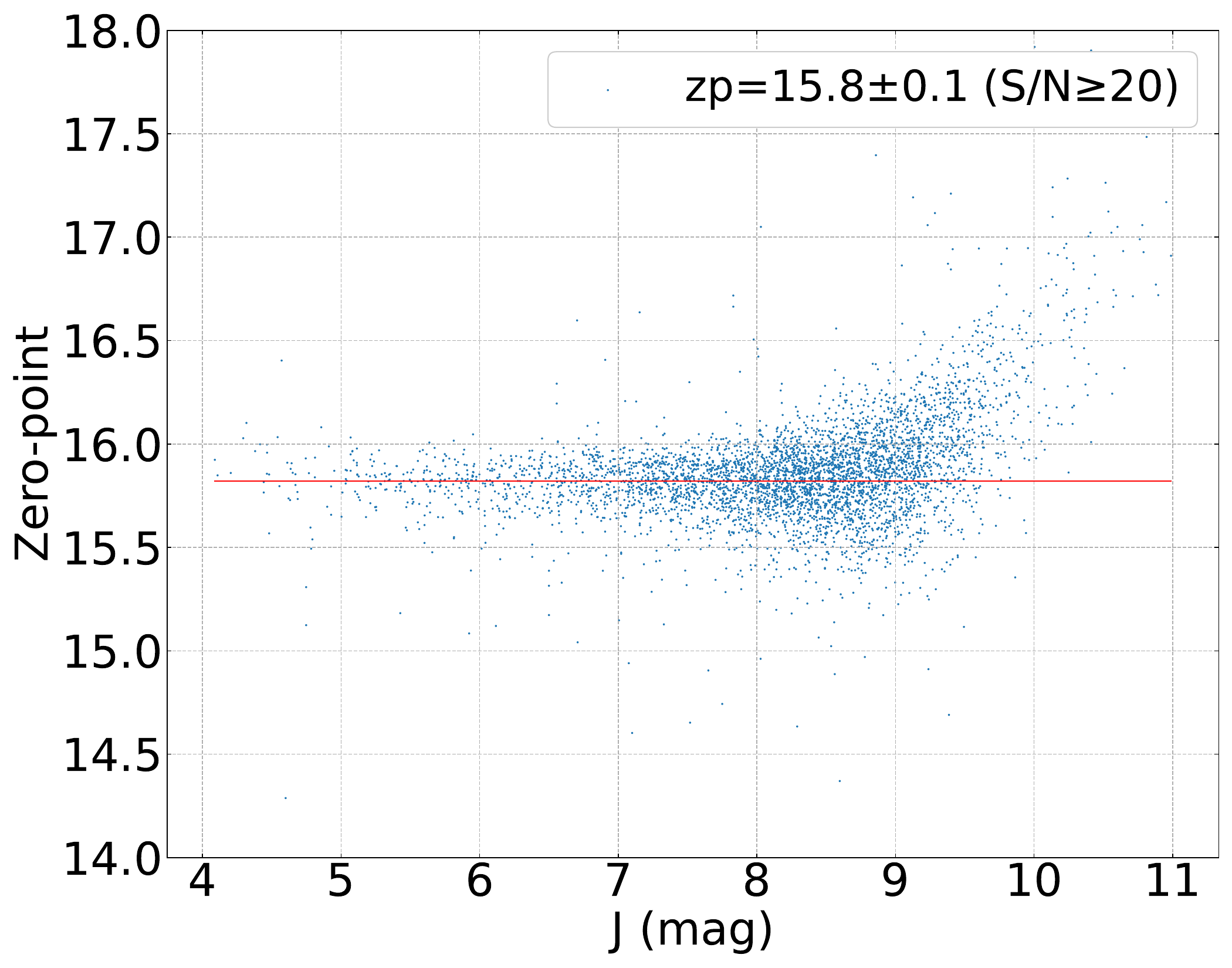}
    \caption{Magnitude zero-points for $J$ (left) and $H$ (right) bands. The analysis is based on sources with S/N\,$\geq$\,20, comprising 6,677 and 4,683 sources for the $J$ and $H$ bands, respectively. The red solid line shows the median values, yielding zero-points of 16.9\,mag and 15.8\,mag for $J$ and $H$ bands, respectively. Outliers may correspond to variable sources.
    }
    \label{fig:zp}
\end{figure*}

Fig.~\ref{fig:zp} shows AIRBT's magnitude zero-points in $J$ and $H$ bands. We determined the system zero-points by calculating the median from sources with S/N $\geq$ 20, yielding 16.9 mag and 15.8 mag for $J$ and $H$ bands, respectively. Consequently, the overall throughput of the system is 19\% and 17\% for the $J$ and $H$ bands, respectively, caused by transmissions of atmosphere, ITO, optics and filter, as well as detector quantum efficiency. Higher overall throughput is expected in Antarctica due to improved atmospheric transmission.

\begin{figure*}
  \centering
  \includegraphics[height=5cm]{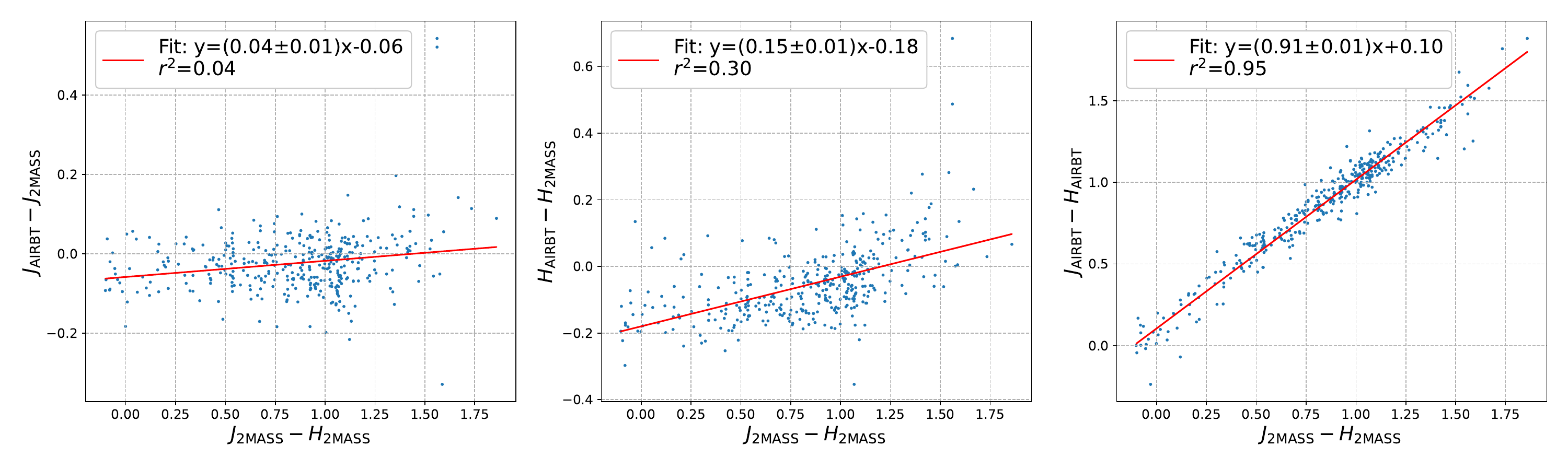}
  \caption{The magnitude transformation between AIRBT and 2MASS for $J$ (left), $H$ (middle), and $J - H$ (right), derived from sources with $S/N \geq 20$. The best linear fit is shown by the red lines.
  }
  \label{fig:color}
\end{figure*}

We compare the magnitude systems of AIRBT and 2MASS in Fig.~\ref{fig:color}. The best linear fit for the magnitude transformation is
\begin{equation}
    J_{\mathrm{AIRBT}} = J_{\mathrm{2MASS}} + 0.05 (J - H)_{\mathrm{2MASS}}
\end{equation}
\begin{equation}
    H_{\mathrm{AIRBT}} = H_{\mathrm{2MASS}} + 0.16 (J - H)_{\mathrm{2MASS}}
\end{equation}
\begin{equation}
    (J - H)_{\mathrm{AIRBT}} = 0.96 (J - H)_{\mathrm{2MASS}} - 0.02
\end{equation}
The $J$ band difference shows marginal color dependence with a linear coefficient of 0.05, while the $H$ band color term reaches 0.16, mainly because the effective $H$ band of AIRBT covers only the blue half of that of 2MASS. 
Additional variations, including telescope, filter, and atmospheric transmission, also contribute to the color dependence. Therefore, applying the magnitude transformation is essential when combining data from the two systems.

\subsection{Photometric precision}
\label{sec:Photometric precision}

\begin{figure*}
    \centering
    \includegraphics[height=6.2cm]{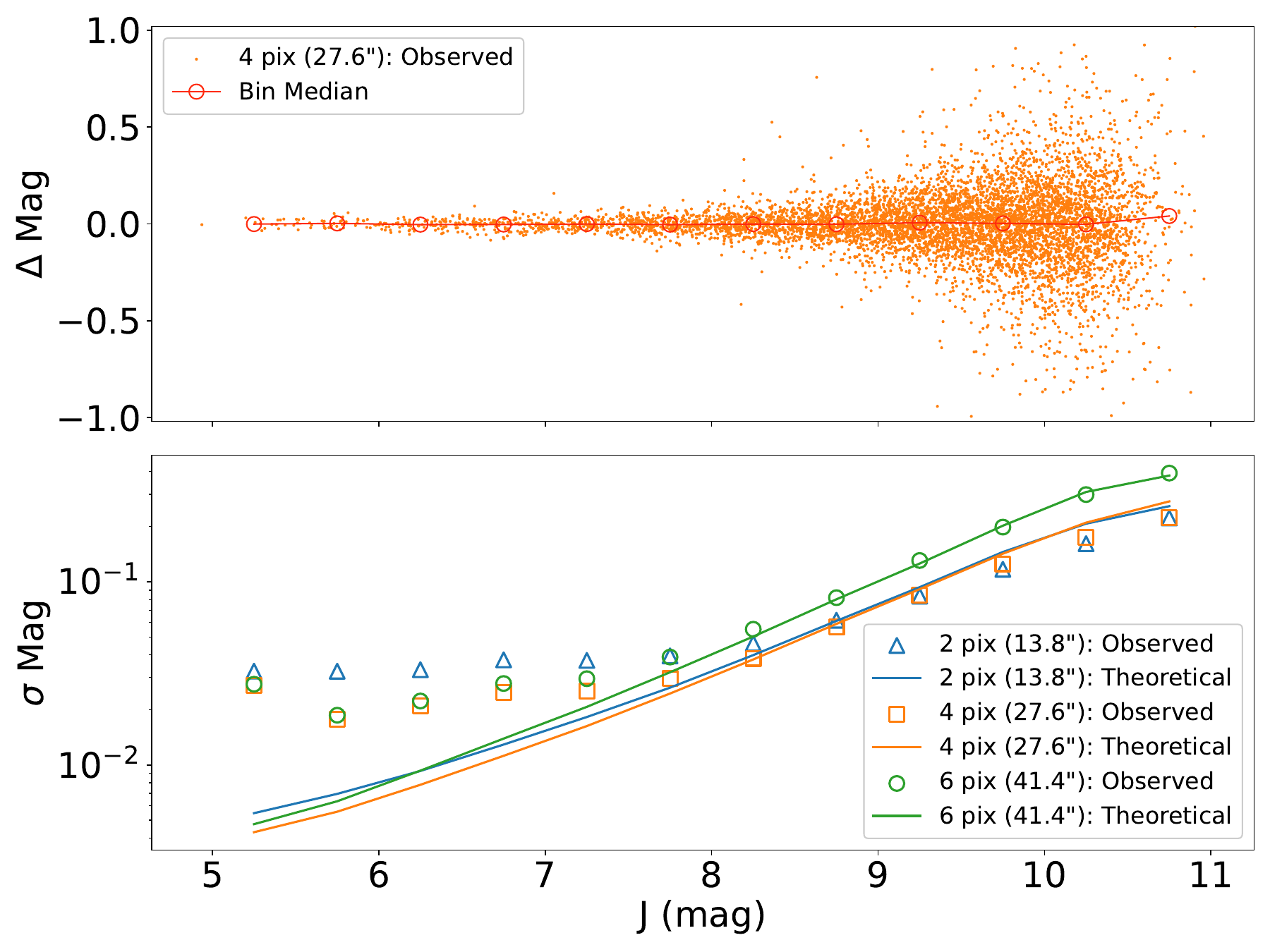}\quad
    \includegraphics[height=6.2cm]{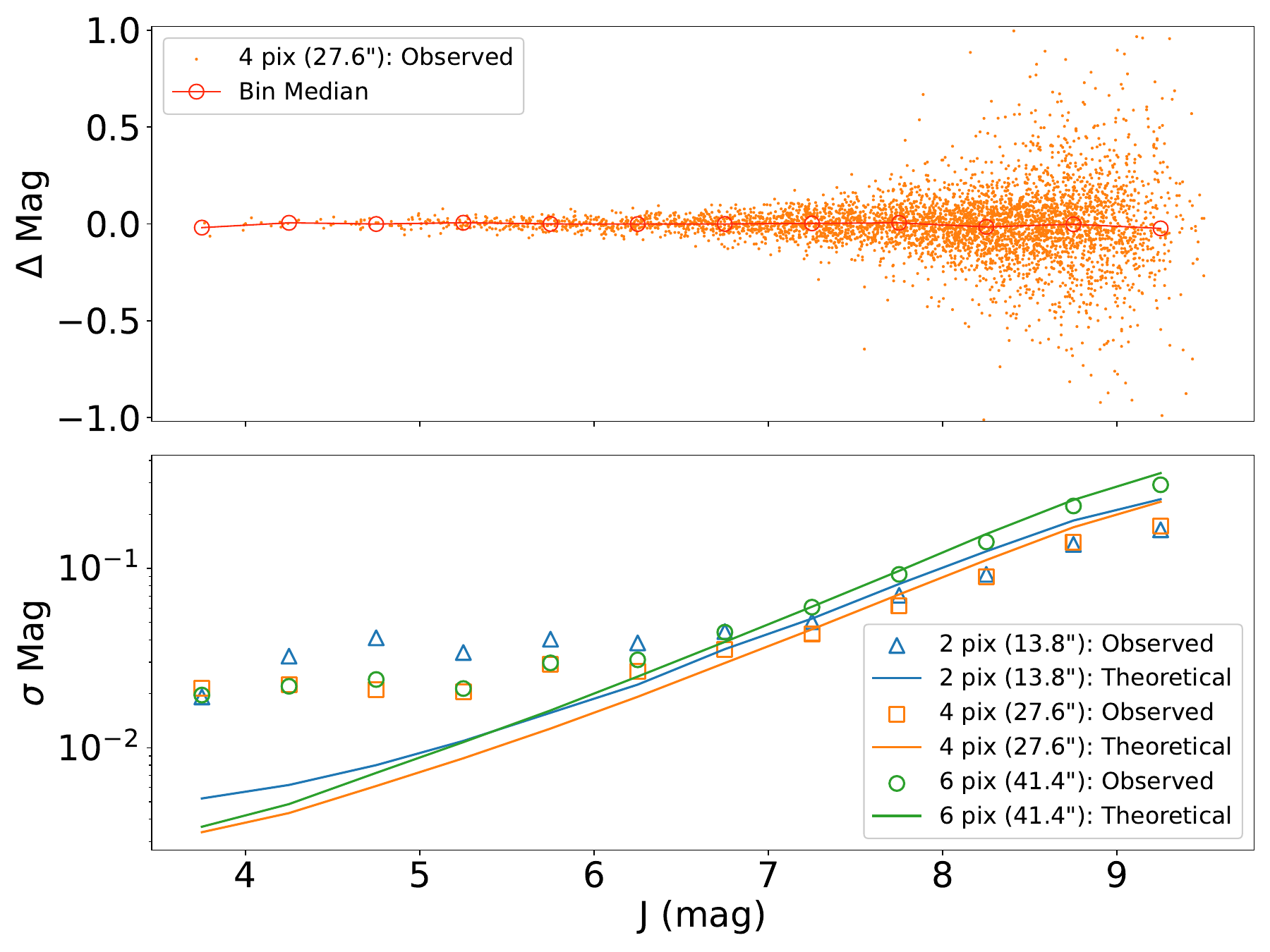}
    \caption{Photometric precision of $J$ (left) and $H$ (right) bands. For each band, the upper panel shows magnitude differences ($\Delta$\,mag) between consecutive exposures, while the lower panel displays photometric errors ($\sigma$\,mag) binned in 0.5-mag intervals for three diameter aperture sizes: 2, 4, and 6 pixels, corresponding to 13.8, 27.6, and 41.4 arcsec, respectively. The best photometric precision reaches 20~mmag, and the limiting magnitude is 11.2 (9.7) in the $J$ ($H$) band.
    }
    \label{fig:dmag-mag}
\end{figure*}

To minimize pixel-related differences, we evaluated photometric precision by comparing two consecutive images with shifts of less than 0.1 pixels. Therefore, the magnitude differences are primarily due to measurement error, which can be calculated as the standard deviation of the former divided by $\sqrt{2}$. We plot the magnitude differences from all image pairs in the upper panels of Fig.~\ref{fig:dmag-mag}, and the practical versus theoretical magnitude errors in the lower panels.

For the faint end, the noise is dominated by sky background and dark current. Smaller apertures provide better precision for faint sources by collecting less noise from the sky background. The 5$\sigma$ limiting magnitude is about 11.2 and 9.7 for the $J$ and $H$ bands, respectively, with a 2-pixel aperture. Both bands are limited in depth by the bright sky and high dark current during the tests.  
The sky background during the tests was measured to be approximately 1,000\,$e^{-}$\,s$^{-1}$\,pixel$^{-1}$ (14.7 mag arcsec\textsuperscript{-2}) in $J$ band and 2,700\,$e^{-}$\,s$^{-1}$\,pixel$^{-1}$ (12.6 mag arcsec\textsuperscript{-2}) in $H$ band. These values are several times brighter than those observed at Dome A, as reported in \citep{2023MNRAS.521.5624Z}. Hence, AIRBT will be able to image deeper at Dome A, benefiting from the dark sky and low dark current.

For bright sources, photon noise becomes dominant. Larger apertures provide better precision for bright sources by collecting more photons, where the noise is dominated by the stellar flux. However, rather than following the ideal photon-noise predictions, the measured precisions for unsaturated stars observed with a 6-pixel aperture approached approximately 0.02 mag. We propose that the noise is dominated by atmospheric scintillation for bright stars.

The approximation of the scintillation variance is given by \citet{1967AJ.....72..747Y}:
\begin{equation}
\sigma_{\text{sc}}^2 = 10\times10^{-6}D^{-3/4}t^{-1}(\cos z)^{-3}\exp(-2h_{\text{obs}}/H),
\end{equation}
where $D$ is the telescope diameter (m), $t$ is the exposure time (s), $h_{\text{obs}}$ is the observatory height (15 m), $z$ is the zenith distance, $H$ is the atmospheric scale height (8000 m), and $\sigma_{\text{sc}}$ is a dimensionless relative error of flux. For our test, this results in a scintillation of about 1\%.
\citet{2015MNRAS.452.1707O} demonstrated that this equation underestimated the median scintillation noise at several major observatories by a factor of about 1.5. This factor would likely increase further during observational tests, due to typically stronger atmospheric turbulence compared to that in world-class observatories. Consequently, scintillation accounts for the predominant photometric noise for bright stars in our test.
By combining six images, the noise could be reduced to about 8 mmag. In particular, the scintillation at Dome A is significantly weaker, making photometric precision at the mmag level achievable for bright stars.

\section{Summary}
\label{sec:Summary}

We have developed AIRBT for simultaneous imaging in $J$ and $H$ bands at Dome A, Antarctica. The goals include investigation of bright variable stars and measurements of the sky brightness in the near-infrared. AIRBT consists of two identical optical tube assemblies, each with a diameter of 15\,cm. Each OTA is equipped with an InGaAs camera with an effective FoV of 1.22\degr\,$\times$\,0.97\degr.

We have tested the InGaAs cameras in detail in the laboratory and found that the middle gain mode is optimal for astronomy. In this mode, the cameras have a readout noise of 93\,$e^-$, a dark current of 605\,$e^-$s\textsuperscript{-1}pixel\textsuperscript{-1}, and a saturation capacity of 146\,ke$^-$. The nonlinearity is better than 1\%, and the fraction of bad pixels is 0.04\%. However, InGaAs cameras exhibit different behavior from CCDs. The exposure time for bias should be at least 10\,\textmu s, rather than using the shortest available one.  There are additional noise increasing with integration time, which may bias the slope of a PTC. A PTC should be obtained by fixing the exposure time and varying the light source to increase the signal, allowing for an unbiased measurement of the gain.

We have evaluated the on-sky performance of AIRBT through observations in Zhuhai. Compared with 2MASS magnitudes, the $J$ magnitude of AIRBT shows only a marginal difference, while the $H$ magnitude shows significant color dependence, as the cutoff wavelength of the InGaAs cameras is only 1.63\,\textmu m. For a 3\,s exposure, the 5$\sigma$ limiting magnitudes are 11.2 mag and 9.7 mag in $J$ and $H$ bands, respectively. The photometric precision is better than 2\% at the bright end, and can be further improved to sub-percent levels by image stacking. Both the depth and precision were limited by the observing environment, including a high dark current from the elevated air temperature, a bright sky background, and strong scintillation. Therefore, the photometric performance has considerable potential for improvement at Dome A.

In January 2023, CHINARE 39 deployed AIRBT at Dome A. They installed AIRBT on one of the equatorial mounts for a KL-DIMM telescope \citep{2020Natur.583..771M} on an 8\,m tower. We will release the data and report the results of sky background measurements and time-domain research in upcoming papers of this series.

\begin{acknowledgements}
B.M. acknowledges the support from the National Key R\&D Program of China (grant No. 2022YFC2807303). AIRBT is financially supported by School of Physics and Astronomy, Sun Yat-sen University.
\end{acknowledgements}

\bibliography{sample701}{}

\begin{thebibliography}{}
\expandafter\ifx\csname natexlab\endcsname\relax\def\natexlab#1{#1}\fi
\providecommand{\url}[1]{\href{#1}{#1}}
\providecommand{\dodoi}[1]{doi:~\href{http://doi.org/#1}{\nolinkurl{#1}}}
\providecommand{\doeprint}[1]{\href{http://ascl.net/#1}{\nolinkurl{http://ascl.net/#1}}}
\providecommand{\doarXiv}[1]{\href{https://arxiv.org/abs/#1}{\nolinkurl{https://arxiv.org/abs/#1}}}

\bibitem[{M.~C.~B. {Ashley} {et~al.}(1996){Ashley}, {Burton}, {Storey},
  {Lloyd}, {Bally}, {Briggs}, \& {Harper}}]{1996PASP..108..721A}
{Ashley}, M. C.~B., {Burton}, M.~G., {Storey}, J. W.~V., {et~al.} 1996,
  \bibinfo{title}{{South Pole Observations of the Near-Infrared Sky
  Brightness},} \pasp, 108, 721, \dodoi{10.1086/133792}

\bibitem[{M.~C.~B. {Ashley} {et~al.}(2010){Ashley}, {Allen}, {Bonner},
  {Bradley}, {Cui}, {Everett}, {Feng}, {Gong}, {Hengst}, {Hu}, {Jiang},
  {Kulesa}, {Lawrence}, {Li}, {Luong-van}, {McCaughrean}, {Moore},
  {Pennypacker}, {Qin}, {Riddle}, {Shang}, {Storey}, {Sun}, {Suntzeff},
  {Tothill}, {Travouillon}, {Walker}, {Wang}, {Yan}, {Yang}, {Yang}, {York},
  {Yuan}, {Zhang}, {Zhang}, {Zhou}, \& {Zhu}}]{2010EAS....40...79A}
{Ashley}, M.~C.~B., {Allen}, G., {Bonner}, C.~S., {et~al.} 2010,
  \bibinfo{title}{{PLATO-a robotic observatory for the Antarctic plateau},} in
  EAS Publications Series, Vol.~40, EAS Publications Series, ed. L.~{Spinoglio}
  \& N.~{Epchtein}, 79--84, \dodoi{10.1051/eas/1040009}

\bibitem[{K. {Batty} {et~al.}(2022){Batty}, {Steele}, \&
  {Copperwheat}}]{2022PASP..134f5001B}
{Batty}, K., {Steele}, I., \& {Copperwheat}, C. 2022,
  \bibinfo{title}{{Laboratory and On-sky Testing of an InGaAs Detector for
  Infrared Imaging},} \pasp, 134, 065001, \dodoi{10.1088/1538-3873/ac71cc}

\bibitem[{E. {Bertin}(2006){Bertin}}]{2006ASPC..351..112B}
{Bertin}, E. 2006, \bibinfo{title}{{Automatic Astrometric and Photometric
  Calibration with SCAMP},} in Astronomical Society of the Pacific Conference
  Series, Vol. 351, Astronomical Data Analysis Software and Systems XV, ed.
  C.~{Gabriel}, C.~{Arviset}, D.~{Ponz}, \& S.~{Enrique}, 112

\bibitem[{E. {Bertin} \& S. {Arnouts}(1996){Bertin} \&
  {Arnouts}}]{1996A&AS..117..393B}
{Bertin}, E., \& {Arnouts}, S. 1996, \bibinfo{title}{{SExtractor: Software for
  source extraction.},} \aaps, 117, 393, \dodoi{10.1051/aas:1996164}

\bibitem[{M. {Birch} {et~al.}(2022){Birch}, {Soon}, {Travouillon}, {Mendel},
  {Taylor}, \& {Tiong}}]{2022JATIS...8a6001B}
{Birch}, M., {Soon}, J., {Travouillon}, T., {et~al.} 2022,
  \bibinfo{title}{{InGaAs focal plane array for transient astronomy in the
  NIR},} Journal of Astronomical Telescopes, Instruments, and Systems, 8,
  016001, \dodoi{10.1117/1.JATIS.8.1.016001}

\bibitem[{C.~S. {Bonner} {et~al.}(2008){Bonner}, {Ashley}, {Lawrence},
  {Storey}, {Luong-Van}, \& {Bradley}}]{2008SPIE.7014E..6IB}
{Bonner}, C.~S., {Ashley}, M. C.~B., {Lawrence}, J.~S., {et~al.} 2008,
  \bibinfo{title}{{Snodar: a new instrument to measure the height of the
  boundary layer on the Antarctic plateau},} in Society of Photo-Optical
  Instrumentation Engineers (SPIE) Conference Series, Vol. 7014, Ground-based
  and Airborne Instrumentation for Astronomy II, ed. I.~S. {McLean} \& M.~M.
  {Casali}, 70146I, \dodoi{10.1117/12.788154}

\bibitem[{C.~S. {Bonner} {et~al.}(2010){Bonner}, {Ashley}, {Cui}, {Feng},
  {Gong}, {Lawrence}, {Luong-Van}, {Shang}, {Storey}, {Wang}, {Yang}, {Yang},
  {Zhou}, \& {Zhu}}]{2010PASP..122.1122B}
{Bonner}, C.~S., {Ashley}, M.~C.~B., {Cui}, X., {et~al.} 2010,
  \bibinfo{title}{{Thickness of the Atmospheric Boundary Layer Above Dome A,
  Antarctica, during 2009},} \pasp, 122, 1122, \dodoi{10.1086/656250}

\bibitem[{M.~G. {Burton} {et~al.}(2000){Burton}, {Ashley}, {Marks},
  {Schinckel}, {Storey}, {Fowler}, {Merrill}, {Sharp}, {Gatley}, {Harper},
  {Loewenstein}, {Mrozek}, {Jackson}, \& {Kraemer}}]{2000ApJ...542..359B}
{Burton}, M.~G., {Ashley}, M.~C.~B., {Marks}, R.~D., {et~al.} 2000,
  \bibinfo{title}{{High-Resolution Imaging of Photodissociation Regions in NGC
  6334},} \apj, 542, 359, \dodoi{10.1086/309510}

\bibitem[{M.~G. {Burton} {et~al.}(2016){Burton}, {Zheng}, {Mould}, {Cooke},
  {Ireland}, {Uddin}, {Zhang}, {Yuan}, {Lawrence}, {Ashley}, {Wu}, {Curtin}, \&
  {Wang}}]{2016PASA...33...47B}
{Burton}, M.~G., {Zheng}, J., {Mould}, J., {et~al.} 2016,
  \bibinfo{title}{{Scientific Goals of the Kunlun Infrared Sky Survey (KISS)},}
  \pasa, 33, e047, \dodoi{10.1017/pasa.2016.38}

\bibitem[{K. {De} {et~al.}(2020){De}, {Hankins}, {Kasliwal}, {Moore}, {Ofek},
  {Adams}, {Ashley}, {Babul}, {Bagdasaryan}, {Burdge}, {Burnham}, {Dekany},
  {Declacroix}, {Galla}, {Greffe}, {Hale}, {Jencson}, {Lau}, {Mahabal},
  {McKenna}, {Sharma}, {Shopbell}, {Smith}, {Soon}, {Sokoloski}, {Soria}, \&
  {Travouillon}}]{2020PASP..132b5001D}
{De}, K., {Hankins}, M.~J., {Kasliwal}, M.~M., {et~al.} 2020,
  \bibinfo{title}{{Palomar Gattini-IR: Survey Overview, Data Processing System,
  On-sky Performance and First Results},} \pasp, 132, 025001,
  \dodoi{10.1088/1538-3873/ab6069}

\bibitem[{Z. {Dong} {et~al.}(2024){Dong}, {Ma}, {Li}, \&
  {Zhang}}]{2024SPIE13103E..1MD}
{Dong}, Z., {Ma}, B., {Li}, J., \& {Zhang}, H. 2024,
  \bibinfo{title}{{Characterizing the noises of InGaAs cameras for astronomical
  observations},} in Society of Photo-Optical Instrumentation Engineers (SPIE)
  Conference Series, Vol. 13103, X-Ray, Optical, and Infrared Detectors for
  Astronomy XI, ed. A.~D. {Holland} \& K.~{Minoglou}, 131031M,
  \dodoi{10.1117/12.3018935}

\bibitem[{G.~A. {Durand} {et~al.}(2014){Durand}, {Tremblin}, {Minier},
  {Reinert}, {Leroy dos Santos}, {Rodriguez}, {Joffrin}, {Busso}, {Tosti},
  {Nucciarelli}, {Dolci}, {Straniero}, {Valentini}, {Abia}, {Christille},
  {Doumayrou}, {Lortholary}, {Charron}, {Lotrus}, {Walter}, {Ronayette},
  {Challita}, {Fromont}, {Condamin}, {Kwon}, \&
  {Tavagnacco}}]{2014SPIE.9145E..0DD}
{Durand}, G.~A., {Tremblin}, P., {Minier}, V., {et~al.} 2014,
  \bibinfo{title}{{Antarctic observations at long wavelengths with the
  IRAIT-ITM Telescope at Dome C},} in Society of Photo-Optical Instrumentation
  Engineers (SPIE) Conference Series, Vol. 9145, Ground-based and Airborne
  Telescopes V, ed. L.~M. {Stepp}, R.~{Gilmozzi}, \& H.~J. {Hall}, 91450D,
  \dodoi{10.1117/12.2056562}

\bibitem[{N. {Earley} {et~al.}(2024){Earley}, {Fucik}, {Fahey}, {Roberts},
  {Smith}, \& {Kasliwal}}]{2024SPIE13096E..3ME}
{Earley}, N., {Fucik}, J., {Fahey}, L., {et~al.} 2024,
  \bibinfo{title}{{Cryoscope Pathfinder: wide-field optical performance
  validation in the lab},} in Society of Photo-Optical Instrumentation
  Engineers (SPIE) Conference Series, Vol. 13096, Ground-based and Airborne
  Instrumentation for Astronomy X, ed. J.~J. {Bryant}, K.~{Motohara}, \&
  J.~R.~D. {Vernet}, 130963M, \dodoi{10.1117/12.3020678}

\bibitem[{N. {Epchtein} {et~al.}(1997){Epchtein}, {de Batz}, {Capoani},
  {Chevallier}, {Copet}, {Fouqu{\'e}}, {Lacombe}, {Le Bertre}, {Pau}, {Rouan},
  {Ruphy}, {Simon}, {Tiph{\`e}ne}, {Burton}, {Bertin}, {Deul}, {Habing},
  {Borsenberger}, {Dennefeld}, {Guglielmo}, {Loup}, {Mamon}, {Ng}, {Omont},
  {Provost}, {Renault}, {Tanguy}, {Kimeswenger}, {Kienel}, {Garzon}, {Persi},
  {Ferrari-Toniolo}, {Robin}, {Paturel}, {Vauglin}, {Forveille}, {Delfosse},
  {Hron}, {Schultheis}, {Appenzeller}, {Wagner}, {Balazs}, {Holl},
  {L{\'e}pine}, {Boscolo}, {Picazzio}, {Duc}, \&
  {Mennessier}}]{1997Msngr..87...27E}
{Epchtein}, N., {de Batz}, B., {Capoani}, L., {et~al.} 1997,
  \bibinfo{title}{{The deep near-infrared southern sky survey (DENIS).},} The
  Messenger, 87, 27

\bibitem[{D. {Frostig} {et~al.}(2022){Frostig}, {Biscoveanu}, {Mo},
  {Karambelkar}, {Dal Canton}, {Chen}, {Kasliwal}, {Katsavounidis}, {Lourie},
  {Simcoe}, \& {Vitale}}]{2022ApJ...926..152F}
{Frostig}, D., {Biscoveanu}, S., {Mo}, G., {et~al.} 2022, \bibinfo{title}{{An
  Infrared Search for Kilonovae with the WINTER Telescope. I. Binary Neutron
  Star Mergers},} \apj, 926, 152, \dodoi{10.3847/1538-4357/ac4508}

\bibitem[{M.~G. {Hidas} {et~al.}(2000){Hidas}, {Burton}, {Chamberlain}, \&
  {Storey}}]{2000PASA...17..260H}
{Hidas}, M.~G., {Burton}, M.~G., {Chamberlain}, M.~A., \& {Storey}, J. W.~V.
  2000, \bibinfo{title}{{Infrared and Submillimetre Observing Conditions on the
  Antarctic Plateau},} \pasa, 17, 260, \dodoi{10.1071/AS00033}

\bibitem[{Y. {Hu} {et~al.}(2014){Hu}, {Shang}, {Ashley}, {Bonner}, {Hu}, {Liu},
  {Li}, {Ma}, {Wang}, \& {Wen}}]{2014PASP..126..868H}
{Hu}, Y., {Shang}, Z., {Ashley}, M. C.~B., {et~al.} 2014,
  \bibinfo{title}{{Meteorological Data for the Astronomical Site at Dome A,
  Antarctica},} \pasp, 126, 868, \dodoi{10.1086/678327}

\bibitem[{Y. {Hu} {et~al.}(2019){Hu}, {Hu}, {Shang}, {Ashley}, {Ma}, {Du},
  {Li}, {Liu}, {Wang}, {Yang}, {Yu}, \& {Zeng}}]{2019PASP..131a5001H}
{Hu}, Y., {Hu}, K., {Shang}, Z., {et~al.} 2019, \bibinfo{title}{{Meteorological
  Data from KLAWS-2G for an Astronomical Site Survey of Dome A, Antarctica},}
  \pasp, 131, 015001, \dodoi{10.1088/1538-3873/aae916}

\bibitem[{M.~M. {Kasliwal} {et~al.}(2025){Kasliwal}, {Earley}, {Smith},
  {Guillot}, {Travouillon}, {Fucik}, {Abe}, {Greffe}, {Agabi}, {Ashley},
  {Triaud}, {Tinyanont}, {Antier}, {Bendjoya}, {Bhattarai}, {Bertz}, {Brugger},
  {Burdanov}, {Caiazzo}, {Carry}, {Casagrande}, {Cenko}, {Cooke}, {De},
  {Dekany}, {Deloupy}, {Dornic}, {Fahey}, {Figer}, {Freeman}, {Frostig},
  {Graham}, {G{\"u}nther}, {Hale}, {Bland-Hawthorn}, {Illuminati}, {Jencson},
  {Karambelkar}, {Key}, {Lau}, {Li}, {Lubin}, {Neill}, {Pahuja}, {Pian}, {de
  Ugarte Postigo}, {Roberts}, {Rodriguez}, {Rose}, {Ruiter}, {Schmider},
  {Simcoe}, {Stein}, {Suarez}, {Taylor}, {Weber}, {Wen}, {de Wit}, {Zarzaca},
  \& {Zimmer}}]{2025PASP..137f5001K}
{Kasliwal}, M.~M., {Earley}, N., {Smith}, R., {et~al.} 2025,
  \bibinfo{title}{{Cryoscope: A Cryogenic Infrared Survey Telescope in
  Antarctica},} \pasp, 137, 065001, \dodoi{10.1088/1538-3873/adc629}

\bibitem[{A. {Lawrence} {et~al.}(2007){Lawrence}, {Warren}, {Almaini}, {Edge},
  {Hambly}, {Jameson}, {Lucas}, {Casali}, {Adamson}, {Dye}, {Emerson},
  {Foucaud}, {Hewett}, {Hirst}, {Hodgkin}, {Irwin}, {Lodieu}, {McMahon},
  {Simpson}, {Smail}, {Mortlock}, \& {Folger}}]{2007MNRAS.379.1599L}
{Lawrence}, A., {Warren}, S.~J., {Almaini}, O., {et~al.} 2007,
  \bibinfo{title}{{The UKIRT Infrared Deep Sky Survey (UKIDSS)},} \mnras, 379,
  1599, \dodoi{10.1111/j.1365-2966.2007.12040.x}

\bibitem[{J.~S. {Lawrence} {et~al.}(2009){Lawrence}, {Ashley}, {Hengst},
  {Luong-van}, {Storey}, {Yang}, {Zhou}, \& {Zhu}}]{2009RScI...80f4501L}
{Lawrence}, J.~S., {Ashley}, M.~C.~B., {Hengst}, S., {et~al.} 2009,
  \bibinfo{title}{{The PLATO Dome A site-testing observatory: Power generation
  and control systems},} Review of Scientific Instruments, 80, 064501,
  \dodoi{10.1063/1.3137081}

\bibitem[{Z.-Y. {Li} {et~al.}(2024){Li}, {Cong}, {Wu}, {Jiang}, {Chen}, {Chen},
  {Li}, {Pei}, {Yao}, {Yang}, {Ji}, \& {Ashley}}]{2024PASP..136k5002L}
{Li}, Z.-Y., {Cong}, J.-N., {Wu}, Z.-X., {et~al.} 2024, \bibinfo{title}{{System
  Design for a Wide Field-of-view Near-infrared Telescope for Dome A in
  Antarctica},} \pasp, 136, 115002, \dodoi{10.1088/1538-3873/ad8d7b}

\bibitem[{N.~P. {Lourie} {et~al.}(2020){Lourie}, {Baker}, {Burruss}, {Egan},
  {F{\.z}r{\'e}sz}, {Frostig}, {Garcia-Zych}, {Ganciu}, {Haworth},
  {Hinrichsen}, {Kasliwal}, {Karambelkar}, {Malonis}, {Simcoe}, \&
  {Zolkower}}]{2020SPIE11447E..9KL}
{Lourie}, N.~P., {Baker}, J.~W., {Burruss}, R.~S., {et~al.} 2020,
  \bibinfo{title}{{The wide-field infrared transient explorer (WINTER)},} in
  Society of Photo-Optical Instrumentation Engineers (SPIE) Conference Series,
  Vol. 11447, Ground-based and Airborne Instrumentation for Astronomy VIII, ed.
  C.~J. {Evans}, J.~J. {Bryant}, \& K.~{Motohara}, 114479K,
  \dodoi{10.1117/12.2561210}

\bibitem[{B. {Ma} {et~al.}(2014){Ma}, {Shang}, {Hu}, {Liu}, {Wang}, \&
  {Wei}}]{2014SPIE.9154E..1TM}
{Ma}, B., {Shang}, Z., {Hu}, Y., {et~al.} 2014, \bibinfo{title}{{A new method
  of CCD dark current correction via extracting the dark Information from
  scientific images},} in Society of Photo-Optical Instrumentation Engineers
  (SPIE) Conference Series, Vol. 9154, High Energy, Optical, and Infrared
  Detectors for Astronomy VI, ed. A.~D. {Holland} \& J.~{Beletic}, 91541T,
  \dodoi{10.1117/12.2055416}

\bibitem[{B. {Ma} {et~al.}(2018){Ma}, {Shang}, {Hu}, {Hu}, {Liu}, {Ashley},
  {Cui}, {Du}, {Fan}, {Feng}, {Huang}, {Gu}, {He}, {Ji}, {Li}, {Li}, {Liu},
  {Tian}, {Tao}, {Wang}, {Wang}, {Wang}, {Wang}, {Wei}, {Wu}, {Xu}, {Yang},
  {Yang}, {Yang}, {Yu}, {Yuan}, {Zhou}, {Zhang}, {Zhang}, {Zhang}, {Zhao},
  {Zhou}, \& {Zhu}}]{2018MNRAS.479..111M}
{Ma}, B., {Shang}, Z., {Hu}, Y., {et~al.} 2018, \bibinfo{title}{{The first
  release of the AST3-1 Point Source Catalogue from Dome A, Antarctica},}
  \mnras, 479, 111, \dodoi{10.1093/mnras/sty1392}

\bibitem[{B. {Ma} {et~al.}(2020){Ma}, {Shang}, {Hu}, {Hu}, {Wang}, {Yang},
  {Ashley}, {Hickson}, \& {Jiang}}]{2020Natur.583..771M}
{Ma}, B., {Shang}, Z., {Hu}, Y., {et~al.} 2020, \bibinfo{title}{{Night-time
  measurements of astronomical seeing at Dome A in Antarctica},} \nat, 583,
  771, \dodoi{10.1038/s41586-020-2489-0}

\bibitem[{D. {Minniti} {et~al.}(2010){Minniti}, {Lucas}, {Emerson}, {Saito},
  {Hempel}, {Pietrukowicz}, {Ahumada}, {Alonso}, {Alonso-Garcia}, {Arias},
  {Bandyopadhyay}, {Barb{\'a}}, {Barbuy}, {Bedin}, {Bica}, {Borissova},
  {Bronfman}, {Carraro}, {Catelan}, {Clari{\'a}}, {Cross}, {de Grijs},
  {D{\'e}k{\'a}ny}, {Drew}, {Fari{\~n}a}, {Feinstein}, {Fern{\'a}ndez
  Laj{\'u}s}, {Gamen}, {Geisler}, {Gieren}, {Goldman}, {Gonzalez}, {Gunthardt},
  {Gurovich}, {Hambly}, {Irwin}, {Ivanov}, {Jord{\'a}n}, {Kerins}, {Kinemuchi},
  {Kurtev}, {L{\'o}pez-Corredoira}, {Maccarone}, {Masetti}, {Merlo},
  {Messineo}, {Mirabel}, {Monaco}, {Morelli}, {Padilla}, {Palma}, {Parisi},
  {Pignata}, {Rejkuba}, {Roman-Lopes}, {Sale}, {Schreiber}, {Schr{\"o}der},
  {Smith}, {Sodr{\'e}}, {Soto}, {Tamura}, {Tappert}, {Thompson}, {Toledo},
  {Zoccali}, \& {Pietrzynski}}]{2010NewA...15..433M}
{Minniti}, D., {Lucas}, P.~W., {Emerson}, J.~P., {et~al.} 2010,
  \bibinfo{title}{{VISTA Variables in the Via Lactea (VVV): The public ESO
  near-IR variability survey of the Milky Way},} \na, 15, 433,
  \dodoi{10.1016/j.newast.2009.12.002}

\bibitem[{A.~M. {Moore} {et~al.}(2016){Moore}, {Kasliwal}, {Gelino}, {Jencson},
  {Jones}, {Kirkpatrick}, {Lau}, {Ofek}, {Petrunin}, {Smith}, {Terebizh},
  {Steinbring}, \& {Yan}}]{2016SPIE.9906E..2CM}
{Moore}, A.~M., {Kasliwal}, M.~M., {Gelino}, C.~R., {et~al.} 2016,
  \bibinfo{title}{{Unveiling the dynamic infrared sky with Gattini-IR},} in
  Society of Photo-Optical Instrumentation Engineers (SPIE) Conference Series,
  Vol. 9906, Ground-based and Airborne Telescopes VI, ed. H.~J. {Hall},
  R.~{Gilmozzi}, \& H.~K. {Marshall}, 99062C, \dodoi{10.1117/12.2233694}

\bibitem[{S. {Murakawa} {et~al.}(2024){Murakawa}, {De}, {Ashley}, {Earley},
  {Hillenbrand}, {Kasliwal}, {Lau}, {Moore}, {Sokoloski}, \&
  {Soria}}]{2024PASP..136j4501M}
{Murakawa}, S., {De}, K., {Ashley}, M. C.~B., {et~al.} 2024,
  \bibinfo{title}{{The First Palomar Gattini-IR Catalog of J-band Light Curves:
  Construction and Public Data Release},} \pasp, 136, 104501,
  \dodoi{10.1088/1538-3873/ad7db1}

\bibitem[{S. {Noll} {et~al.}(2012){Noll}, {Kausch}, {Barden}, {Jones},
  {Szyszka}, {Kimeswenger}, \& {Vinther}}]{2012A&A...543A..92N}
{Noll}, S., {Kausch}, W., {Barden}, M., {et~al.} 2012, \bibinfo{title}{{An
  atmospheric radiation model for Cerro Paranal. I. The optical spectral
  range},} \aap, 543, A92, \dodoi{10.1051/0004-6361/201219040}

\bibitem[{J. {Osborn} {et~al.}(2015){Osborn}, {F{\"o}hring}, {Dhillon}, \&
  {Wilson}}]{2015MNRAS.452.1707O}
{Osborn}, J., {F{\"o}hring}, D., {Dhillon}, V.~S., \& {Wilson}, R.~W. 2015,
  \bibinfo{title}{{Atmospheric scintillation in astronomical photometry},}
  \mnras, 452, 1707, \dodoi{10.1093/mnras/stv1400}

\bibitem[{P.~P. {Pedersen} {et~al.}(2024){Pedersen}, {Queloz}, {Garcia},
  {Schacke}, {Delrez}, {Demory}, {Ducrot}, {Dransfield}, {Gillon}, {Hooton},
  {Jan{\'o}-Mu{\~n}oz}, {Jehin}, {Sebastian}, {Timmermans}, {Thompson},
  {Triaud}, {de Wit}, \& {Z{\'u}{\~n}iga-Fern{\'a}ndez}}]{2024SPIE13096E..3XP}
{Pedersen}, P.~P., {Queloz}, D., {Garcia}, L., {et~al.} 2024,
  \bibinfo{title}{{Infrared photometry with InGaAs detectors: first light with
  SPECULOOS},} in Society of Photo-Optical Instrumentation Engineers (SPIE)
  Conference Series, Vol. 13096, Ground-based and Airborne Instrumentation for
  Astronomy X, ed. J.~J. {Bryant}, K.~{Motohara}, \& J.~R.~D. {Vernet},
  130963X, \dodoi{10.1117/12.3018320}

\bibitem[{A. {Phillips} {et~al.}(1999){Phillips}, {Burton}, {Ashley}, {Storey},
  {Lloyd}, {Harper}, \& {Bally}}]{1999ApJ...527.1009P}
{Phillips}, A., {Burton}, M.~G., {Ashley}, M.~C.~B., {et~al.} 1999,
  \bibinfo{title}{{The Near-Infrared Sky Emission at the South Pole in
  Winter},} \apj, 527, 1009, \dodoi{10.1086/308097}

\bibitem[{Z. {Shang}(2020){Shang}}]{2020RAA....20..168S}
{Shang}, Z. 2020, \bibinfo{title}{{Astronomy from Dome A in Antarctica},}
  Research in Astronomy and Astrophysics, 20, 168,
  \dodoi{10.1088/1674-4527/20/10/168}

\bibitem[{S.-C. {Shi} {et~al.}(2016){Shi}, {Paine}, {Yao}, {Lin}, {Li}, {Duan},
  {Matsuo}, {Zhang}, {Yang}, {Ashley}, {Shang}, \& {Hu}}]{2016NatAs...1E...1S}
{Shi}, S.-C., {Paine}, S., {Yao}, Q.-J., {et~al.} 2016,
  \bibinfo{title}{{Terahertz and far-infrared windows opened at Dome A in
  Antarctica},} Nature Astronomy, 1, 0001, \dodoi{10.1038/s41550-016-0001}

\bibitem[{R.~A. {Simcoe} {et~al.}(2019){Simcoe}, {F{\H{u}}r{\'e}sz},
  {Sullivan}, {Hellickson}, {Malonis}, {Kasliwal}, {Shectman}, {Kollmeier}, \&
  {Moore}}]{2019AJ....157...46S}
{Simcoe}, R.~A., {F{\H{u}}r{\'e}sz}, G., {Sullivan}, P.~W., {et~al.} 2019,
  \bibinfo{title}{{Background-limited Imaging in the Near Infrared with Warm
  InGaAs Sensors: Applications for Time-domain Astronomy},} \aj, 157, 46,
  \dodoi{10.3847/1538-3881/aae094}

\bibitem[{G. {Sims} {et~al.}(2012){Sims}, {Ashley}, {Cui}, {Everett}, {Feng},
  {Gong}, {Hengst}, {Hu}, {Kulesa}, {Lawrence}, {Luong-Van}, {Ricaud}, {Shang},
  {Storey}, {Wang}, {Yang}, {Yang}, {Zhou}, \& {Zhu}}]{2012PASP..124...74S}
{Sims}, G., {Ashley}, M. C.~B., {Cui}, X., {et~al.} 2012,
  \bibinfo{title}{{Precipitable Water Vapor above Dome A, Antarctica,
  Determined from Diffuse Optical Sky Spectra},} \pasp, 124, 74,
  \dodoi{10.1086/664077}

\bibitem[{M.~F. {Skrutskie} {et~al.}(2006){Skrutskie}, {Cutri}, {Stiening},
  {Weinberg}, {Schneider}, {Carpenter}, {Beichman}, {Capps}, {Chester},
  {Elias}, {Huchra}, {Liebert}, {Lonsdale}, {Monet}, {Price}, {Seitzer},
  {Jarrett}, {Kirkpatrick}, {Gizis}, {Howard}, {Evans}, {Fowler}, {Fullmer},
  {Hurt}, {Light}, {Kopan}, {Marsh}, {McCallon}, {Tam}, {Van Dyk}, \&
  {Wheelock}}]{2006AJ....131.1163S}
{Skrutskie}, M.~F., {Cutri}, R.~M., {Stiening}, R., {et~al.} 2006,
  \bibinfo{title}{{The Two Micron All Sky Survey (2MASS)},} \aj, 131, 1163,
  \dodoi{10.1086/498708}

\bibitem[{J. {Soon} {et~al.}(2020){Soon}, {Adams}, {De}, {Galla}, {Hankins},
  {Kasliwal}, {Moore}, {Adams}, {Antoszewski}, {Ashley}, {Babul},
  {Bland-Hawthorn}, {Cooke}, {De Marco}, {Delacroix}, {Devillepoix}, {Ellis},
  {Freeman}, {Hale}, {Heger}, {Jencson}, {Lau}, {McKenna}, {Ofek}, {Ryder},
  {Simcoe}, {Sokoloski}, {Soria}, {Smith}, \&
  {Travouillon}}]{2020SPIE11203E..07S}
{Soon}, J., {Adams}, D., {De}, K., {et~al.} 2020, \bibinfo{title}{{Wide-field
  dynamic astronomy in the near-infrared with Palomar Gattini-IR and DREAMS},}
  in Society of Photo-Optical Instrumentation Engineers (SPIE) Conference
  Series, Vol. 11203, Advances in Optical Astronomical Instrumentation 2019,
  ed. S.~C. {Ellis} \& C.~{d'Orgeville}, 1120307, \dodoi{10.1117/12.2539594}

\bibitem[{R. {Strausbaugh} {et~al.}(2018){Strausbaugh}, {Jackson}, \&
  {Butler}}]{2018PASP..130i5001S}
{Strausbaugh}, R., {Jackson}, R., \& {Butler}, N. 2018, \bibinfo{title}{{Night
  Vision for Small Telescopes},} \pasp, 130, 095001,
  \dodoi{10.1088/1538-3873/aaca2a}

\bibitem[{P.~W. {Sullivan} {et~al.}(2013){Sullivan}, {Croll}, \&
  {Simcoe}}]{2013PASP..125.1021S}
{Sullivan}, P.~W., {Croll}, B., \& {Simcoe}, R.~A. 2013,
  \bibinfo{title}{{Precision of a Low-Cost InGaAs Detector for Near Infrared
  Photometry},} \pasp, 125, 1021, \dodoi{10.1086/672573}

\bibitem[{W. {Sutherland} {et~al.}(2015){Sutherland}, {Emerson}, {Dalton},
  {Atad-Ettedgui}, {Beard}, {Bennett}, {Bezawada}, {Born}, {Caldwell}, {Clark},
  {Craig}, {Henry}, {Jeffers}, {Little}, {McPherson}, {Murray}, {Stewart},
  {Stobie}, {Terrett}, {Ward}, {Whalley}, \& {Woodhouse}}]{2015A&A...575A..25S}
{Sutherland}, W., {Emerson}, J., {Dalton}, G., {et~al.} 2015,
  \bibinfo{title}{{The Visible and Infrared Survey Telescope for Astronomy
  (VISTA): Design, technical overview, and performance},} \aap, 575, A25,
  \dodoi{10.1051/0004-6361/201424973}

\bibitem[{G. {Tosti} {et~al.}(2006){Tosti}, {Busso}, {Nucciarelli}, {Bagaglia},
  {Roncella}, {Mancini}, {Castellini}, {Mariotti}, {Babucci}, {Chiocci},
  {Straniero}, {Dolci}, {Valentini}, {di Varano}, {Pelusi}, {Di Rico}, {Ragni},
  {Abia}, {Dom{\'\i}nguez}, {Corcione}, {Porcu}, {Conconi}, {De Caprio},
  {Riva}, {Molinari}, {Zerbi}, {Bortoletto}, {Bonoli}, {D'Alessandro},
  {Colom{\'e}}, {Isern}, {Briguglio}, {Cacciani}, {Farnesini}, {Checcucci}, \&
  {Strassmeier}}]{2006SPIE.6267E..1HT}
{Tosti}, G., {Busso}, M., {Nucciarelli}, G., {et~al.} 2006,
  \bibinfo{title}{{The International Robotic Antarctic Infrared Telescope
  (IRAIT)},} in Society of Photo-Optical Instrumentation Engineers (SPIE)
  Conference Series, Vol. 6267, Ground-based and Airborne Telescopes, ed. L.~M.
  {Stepp}, 62671H, \dodoi{10.1117/12.670302}

\bibitem[{C. {Yang} {et~al.}(2025){Yang}, {Ji}, {Li}, {Cong}, {Chen}, {Ashley},
  {Wu}, {Li}, {Luo}, {Xi}, {Zhou}, {Zhao}, {Zhou}, {Hao}, {Zhang}, {Zhou}, \&
  {Jiang}}]{2025AJ....169..228Y}
{Yang}, C., {Ji}, T., {Li}, Z., {et~al.} 2025, \bibinfo{title}{{First Daytime
  Near-infrared Photometric Observations at Antarctic Dome A},} \aj, 169, 228,
  \dodoi{10.3847/1538-3881/adbd0e}

\bibitem[{X. {Yang} {et~al.}(2021){Yang}, {Shang}, {Hu}, {Hu}, {Ma}, {Wang},
  {Cao}, {Ashley}, \& {Wang}}]{2021MNRAS.501.3614Y}
{Yang}, X., {Shang}, Z., {Hu}, K., {et~al.} 2021, \bibinfo{title}{{Cloud cover
  and aurora contamination at dome A in 2017 from KLCAM},} \mnras, 501, 3614,
  \dodoi{10.1093/mnras/staa3824}

\bibitem[{A.~T. {Young}(1967){Young}}]{1967AJ.....72..747Y}
{Young}, A.~T. 1967, \bibinfo{title}{{Photometric error analysis. VI.
  Confirmation of Reiger's theory of scintillation},} \aj, 72, 747,
  \dodoi{10.1086/110303}

\bibitem[{C. {Yu} {et~al.}(2017){Yu}, {Li}, {Yang}, {Huang}, {Shao}, {Zhang},
  \& {Gong}}]{2017InPhT..85...74Y}
{Yu}, C., {Li}, X., {Yang}, B., {et~al.} 2017, \bibinfo{title}{{Noise
  characteristics analysis of short wave infrared InGaAs focal plane arrays},}
  Infrared Physics and Technology, 85, 74,
  \dodoi{10.1016/j.infrared.2017.05.017}

\bibitem[{J. {Zhang} {et~al.}(2023){Zhang}, {Zhang}, {Tang}, {Wang}, {Jiang},
  {Ashley}, {Ji}, {Zhang}, {Feng}, {Wang}, {Zeng}, {Zhang}, {Chen}, {Chen},
  {Jia}, {Zhang}, {Zhou}, {hu}, {Wang}, \& {Zhu}}]{2023MNRAS.521.5624Z}
{Zhang}, J., {Zhang}, Y.-h., {Tang}, Q.-J., {et~al.} 2023,
  \bibinfo{title}{{Sky-brightness measurements in J, H, and Ks bands at DOME A
  with NISBM and early results},} \mnras, 521, 5624,
  \dodoi{10.1093/mnras/stad775}

\bibitem[{H. {Zou} {et~al.}(2010){Zou}, {Zhou}, {Jiang}, {Ashley}, {Cui},
  {Feng}, {Gong}, {Hu}, {Kulesa}, {Lawrence}, {Liu}, {Luong-Van}, {Ma},
  {Moore}, {Pennypacker}, {Qin}, {Shang}, {Storey}, {Sun}, {Travouillon},
  {Walker}, {Wang}, {Wang}, {Wu}, {Wu}, {Xia}, {Yan}, {Yang}, {Yang}, {Yao},
  {Yuan}, {York}, {Zhang}, \& {Zhu}}]{2010AJ....140..602Z}
{Zou}, H., {Zhou}, X., {Jiang}, Z., {et~al.} 2010, \bibinfo{title}{{Sky
  Brightness and Transparency in the i-band at Dome A, Antarctica},} \aj, 140,
  602, \dodoi{10.1088/0004-6256/140/2/602}

\end{thebibliography}
\bibliographystyle{aasjournalv7}



\end{document}